RUQIAN LU

# KNOWWARE
the third star
after Hardware
and Software

Polimetrica®

PUBLISHING STUDIES

directed by Giandomenico Sica

VOLUME 1

Work supported by:
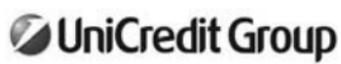

**RUQIAN LU**

# **KNOWWARE**
**the third star
after Hardware
and Software**

Polimetrica

**Note for the Reader**

In our view, doing research means building new knowledge, setting new questions, trying to find new answers, assembling and dismantling frames of interpretation of reality.

**Do you want to participate actively in our research activities?**

**Submit new questions!**

Send an email to the address **questions@polimetrica.org** and include in the message your list of questions related to the subject of this book.

Your questions can be published in the next edition of the book, together with the author's answers.

**Please do it.**

**This operation only takes you a few minutes but it is very important for us, in order to develop the contents of this research.**

Thank you very much for your help and cooperation!

We're open to discuss further collaborations and proposals.
If you have any idea, please contact us at the following address:

*Editorial office*
*POLIMETRICA*
*Corso Milano 26*
*20052 Monza MI Italy*
*Phone: ++39.039.2301829*
*E-mail: info@polimetrica.org*

**We are looking forward to getting in touch with you.**

# L<small>IST OF</small> Q<small>UESTIONS</small>

# **Preface**

As pointed out by Prof. Feigenbaum, we are entering an era of "knowledge industry in which knowledge itself will be a salable commodity like food and oil. Knowledge itself is to become the new wealth of nations."[9] Every commodity, including physical and non physical, has its own form. It is interesting to ask the question: "which form should be taken by knowledge as a commodity?" The answer varies from case to case. In case of industrial manufacturing, it is the patents. In case of printed or electronic publication, it is the copy rights. However, when the case of software is considered, we have found that, much to our surprise, there is no explicit form of (the) knowledge (contained in it) as a commodity, except the software itself. But for software it is very difficult to apply for a patent. One usually assigns copy rights to software. Copy rights do not protect the knowledge content of software. To find an appropriate form of knowledge as a commodity in case when it is contained in software, we must separate knowledge from software codes and assign a different form of intellectual property to knowledge than to software.

   Traditional forms of knowledge organization, like knowledge base, are not suitable for this task. Firstly, a knowledge base is usually contained in a larger software module called

"knowledge based system". The both are tightly connected with a specifically tailored interface. It is not easy to separate a knowledge base from the embracing knowledge based system and to use it in another environment. In this sense the knowledge base is context sensitive and is not an independent commodity. Secondly, a knowledge base is often attached with a management system, such that it is still a mix of knowledge and software codes even without considering the embracing knowledge based system. This management system for the knowledge base can be made arbitrary large and complex, such that the independence of knowledge base becomes a problem again.

We believe that we have found an appropriate form of commodity for the knowledge contained in software. This form is knowware. The idea of knowware was born in a long term project of automatic generation of knowledge based systems, which we have been undertaking since more than 15 years ago. A major result of this long term research was that the development of knowledge based systems can be divided in two parts: the development of a relative stable system shell and that of a movable X knowledge base. These two parts can be developed separately and independently, even by different teams without communication between them. It is only required that some rigorous standard has been worked out before and is strictly followed during the whole process of development. Since we have developed a technique called pseudo natural language understanding, the development of X knowledge base was made automatic. A consequence of this result was that the development of the whole knowledge based system was also made automatic to a large extent, because the system shell is relatively stable. The key difference between an X knowledge base and a traditional knowledge base is its modularity, independence of development and interchangeability.

Seeing that separating knowledge base development from system shell development brings great benefits and advantages, we have generalized the applicability of above idea to more general application software where there is no explicit knowledge base. We request that the otherwise distributed knowledge in such software should be concentrated to form a knowledge core, which should be as modular, independent and interchangeable as the X knowledge base. The second generalization of the above idea was to promote the knowledge core concept one step further to make it an independent commodity. As an industrial commodity it has to meet some standard and is thus context insensitive. We call this commodity knowware. A knowware should have all properties a usual commodity has. The most important meaning of knowware is that it plays an equal significant role in IT just as hardware and software play. We predict that the research and development of knowware will become an independent research interest of computer science and suggest that the IT industry should move from a dipole world (hardware and software) to a triple pole world (hardware, software and knowware).

I thank Dott. Giandomenico Sica and Polimetrica a lot for inviting me to write this book. By this invitation I have got a new chance of "advertising" my idea about knowware. This book is not just a summary of our past work. It contains significant improvement and further development of the knowware concept. I hope it will attract more attention from colleagues and friends of computer science community.

Ruqian Lu
Institute of Mathematics
Academy of Mathematics and System Sciences
Chinese Academy of Sciences
On the day of Chinese Mid-Autumn Festival 2007

*What is knowware?*

**KEYWORDS:**
KNOWWARE, SOFTWARE, HARDWARE, MIXWARE

Currently, we define a knowware as the computer representation of a read only knowledge module that is independent, commercialized, suitable for computer manipulation, meeting some industrial standard, equipped with detailed documentation and embeddable in software or hardware for use.

In more details: independence means the production, sale and use of a knowware is independent of any particular hardware or software; commercialized means knowware is a commodity to be sold on the market; suitable for computer manipulation means knowware is implemented in some (high level or low level) code, which can be interpreted by a computer program; meeting some industrial standard means that all input-output interfaces of a knowware should follow some national or international standard; equipped with detailed documentation means there should be a detailed specification on the functions, the way of use and maintenance, the interface standard, the version number, etc. of the knowware, including a license or patent declaration; embeddable in software or hardware means knowware can be provided in form of macro code (represented in some media) or micro code (represented in hardware circuits).

A knowware is also called a knowware module if it is embedded in some software or hardware. The software or hardware with imbedded knowware modules is called a mixware. Nevertheless, mixware can still be called software or hardware if no ambiguity is raised. In the context of this book, mixware does not involve hardware.

*Are there some typical examples of knowware?*

**KEYWORDS:**
KNOWWARE EXAMPLE, TAX MANAGEMENT SOFTWARE, ICAI, PAPER SEARCH ENGINE, DOLPHIN, AOYING ZHOU

Knowware is a new idea. Therefore we cannot yet buy a knowware for you from the market. However we can show you many examples that are not yet knowware in its perfect form, but already have some characteristic features of knowware. Imagine a MP3 player, which is a combination of hardware and software. All the songs collected in it can be considered as knowledge that forms the basis for the MP3 player to play music. If we wrap a set of songs together in some standard representation and commercialized form, then this package of songs may be considered as simplest knowware.

The second example is tax management software. The tax regulations (considered as knowledge used by tax calculation) published by the government and operated by this tax management software may form a knowware. Each time when the tax law is changed, one does not have to renew the whole tax management software. One just needs to replace the knowware containing old tax law by a knowware containing the new tax law. As a matter of fact, we expect that the governmental tax departments will publish their new tax law in two forms at the same time: a paper version for human reading and a knowware version for computer use. All that the system manager has to do is buying this new knowware and installing it on their tax management software.

The third example is an ICAI system, which contains a course material, say for mathematics education. According to our principle, this course material should be organized as a knowware. Assume a learner wants to study mathematics in his spare time. He starts from learning the junior class mathematics course. For that purpose, he buys an ICAI software package including a junior class mathematics knowware. One year later, the same person wants to improve himself further by studying senior class mathematics course. He needs only to replace the knowware containing junior class mathematics knowledge with one containing senior class mathematics knowledge. He does not need to buy a new ICAI package as a whole.

The fourth example is an ad hoc paper search engine, called Dolphin, developed by the group of Aoying Zhou at Fudan University, Shanghai Key Lab of intelligent information processing[38]. This search engine collects papers from the World Wide Web based on user instructions. It makes use of focused crawlers to locate such papers and extracts basic information from them. Dolphin has already collected 0.76 million computer science papers published in the period 1970-2004. The collected paper information is made to a data warehouse. Dolphin provides a set of tools for analyzing this information. For example, by inputting the keyword "data mining", which denotes a subfield of computer science, one can ask Dolphin to analyze relevant data and to provide results of the analysis, such as annual number of published papers, development tendencies of hot research points of data mining, etc. This data warehouse can be considered as a knowware in preliminary form. The tool set then can be considered as knowledge middleware of the kind "knowledge analysis". Figure 1 illustrates the structure of the Dolphin crawler. Figure 2 is a sample diagram showing the results of knowledge analysis on the hot topics of data mining.

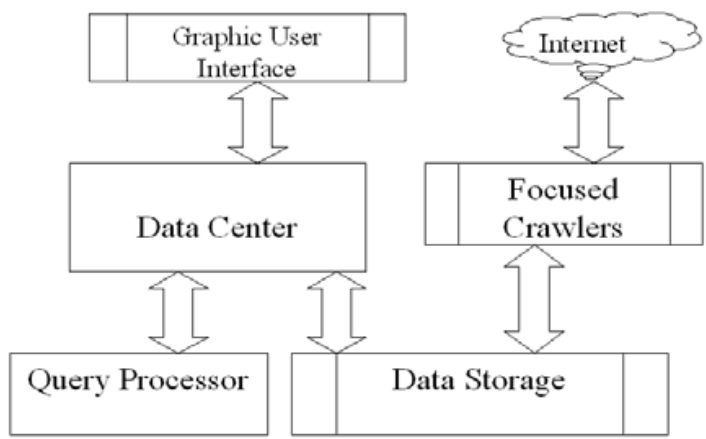

Figure 1. System architecture of Dolphin[38]

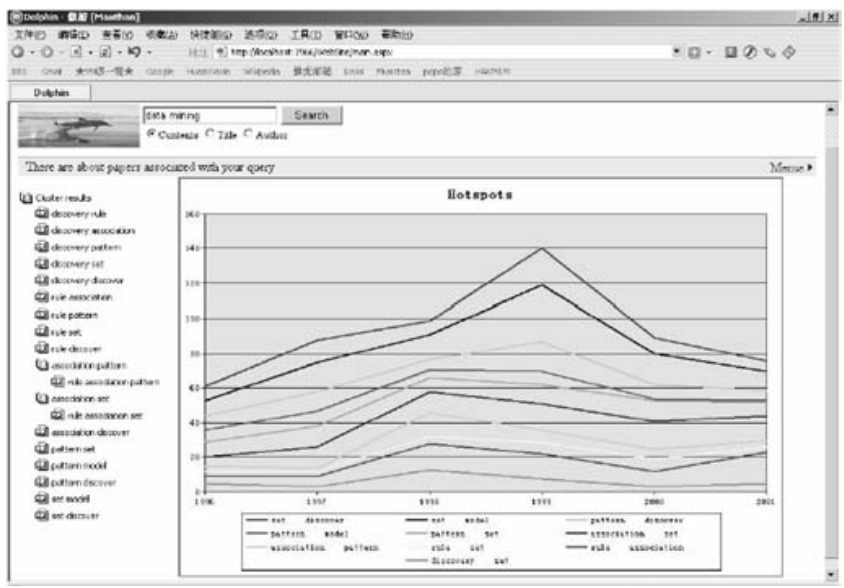

Figure 2. Development tendency of hotspots
of data mining 1994-2003[38] (Dolphin Report)

*If software means knowledge, then software is itself knowware, why do we need another concept of knowware?*

**KEYWORDS:**
APPLICATION KNOWLEDGE, SUPPORT KNOWLEDGE, ACTIVE KNOWLEDGE, FROZEN KNOWLEDGE, COMMON KNOWLEDGE, INDIVIDUAL KNOWLEDGE

Once when the author was talking about knowware at an academic conference, someone of the audience has posed the question: "Software itself is a block of knowledge, how could you separate just a part of software and call it knowledge?" This was really a good question. This listener was right. The late prof. Xiwen Ma has pointed out that software is condensation and crystallization of knowledge. That means the whole piece of software is knowledge. The author of this book agrees with this conclusion.

This point of view can be further generalized. Hardware is also knowledge, because each computer has an instruction set and circuits for implementing the instructions. So computer can be considered as a block of knowledge in its hardware form. Computer network is also knowledge, because each network is based on a set of protocols and firewalls. It may even have a network operating system.

If everything of software and hardware is knowledge, then how can we separate a part of it and say that only this part of knowledge should be made to knowware, and all other parts shouldn't? The key of the answer to this question is to consider the pragmatic aspects of the knowledge. Take a computer as an example. Its instruction set is used to support the running of programs. Without the instruction set it is impossible for any program to run. But on the other hand, the instruction set would be meaningless if there were no programs running on it. That means the meaning of the instruction set is only in its supporting the program running. In this sense, we differentiate the knowledge embodied in some (hardware or software) device in two types: application knowledge and support knowledge. The former is the knowledge that serves the users directly. The latter exists only in background and is used to support the application of the former. The instruction set is just the support knowledge of a computer, whose application knowledge, the program, is usually separated from it and interchangeable. However, the case of software looks different. Its application knowledge and support knowledge are mixed together. What we want to do is just separate application knowledge of software from its support knowledge.

Another dimension of differentiating knowledge contained in software is its stability. Stability of knowledge means that any piece of software embodying this knowledge is not changed frequently. Thus all system software, including operating systems, is stable. Vary rarely a system software is replaced by a new one. On the other hand, the application programs change frequently. Each program gives way to a next one when its run is finished. They are not stable. Even within a program one may find parts more stable and parts less stable. In this sense we can also call the knowledge contained in stable programs or program parts frozen knowledge in contrast to active knowledge that is contained in non stable programs or program parts. It is just active knowledge that is suitable for producing knowware.

Since knowware is a commodity, each time when we want to produce a knowware we should be aware of the benefit/cost relation of the knowware to be produced. Not every piece of knowledge is worthwhile to be coined in form of knowware. For example, a piece of knowledge that is useful in the software currently under development, but hardly useful for other software is not worthwhile to be coined as knowware.

This principle leads to another classification: common knowledge and individual knowledge. Generally speaking, purely individual knowledge is not suitable for knowware production, unless there is special reason to do that.

Conclusion: not every part of software needs to be made to knowware.

*What is the most essential difference between knowware and software?*

**KEYWORDS:**
APPLICATION KNOWWARE, SYSTEM KNOWWARE

Table 1 summarizes the major differences between software (in the traditional sense) and knowware.

|  | **Conventional Software** | **Knowware** |
|---|---|---|
| Technique Content | both domain knowledge and software techniques | only domain knowledge |
| Life Cycle | mainly determined by variations of customer requirement | mainly determined by accumulations and evolution of knowledge |
| Classification | normally can be divided into system software and application software | consists of application knowware only, "system knowware" is a software |
| Developer | primly only software engineers (sometimes supported by domain experts) | experts from various domains (sometimes supported by knowledge engineers) |
| Intellectual Property | software copyright | knowledge patent |

Table 1. Major Differences between Knowware and Software

There is one more basic difference between software and knowware: software, in particular application software, is almost always goal oriented. All software is developed to solve some specified problem. However knowware is usually not limited to any specific application. It is universal purpose oriented. For example, a knowware containing chemical knowledge can serve any application making use of knowledge about chemistry. It may be used as a part of student's chemistry course, or a part of some encyclopedia, or a knowledge base for some drug production factory, etc. In a word, any knowware is produced to solve problems where it can. In this sense, software is in its essence problem oriented, while knowware is in its essence service oriented.

*Is there a meaningful classification of knowware?*

**KEYWORDS:**
PUBLIC KNOWWARE, PRIVATE KNOWWARE, KNOWWARE OUTSOURCING

We have explained above that knowware is not classified in system knowware and application knowware. All knowware is application oriented. But we have another way to classify knowware, namely to classify them in public knowware and private knowware. Public knowware contains public knowledge, which can be acquired and used by everybody. Private knowware contains only private knowledge that is not open to the public and whose use is limited to a small group of people. Therefore you can only buy public knowware from the open market, while private knowware is only obtainable within a small circle of people or institutions.

All examples we have given above are public knowware. Some examples of private knowware are listed below. In an ERP or CRM system, that part representing the business intelligence may be made to private knowware. It contains rules, strategies and heuristics of managing a business. This kind of knowledge is highly confidential and should be kept secrete. For a large institution like a big hospital or a big university, the management information system is operating on a big data base or a set of big data bases. The information contained in it is not useful for anybody outside of the hospital or university. It can be made to a private knowware, either.

One would ask the question that since these knowware are private, their development must be also a private activity. Then how can they be a commodity? In order to get an answer to this question, one just has to look at the general tendency of commercializing software development. Nowadays, many big institutions do not develop software themselves, but order it in the way of software outsourcing. This idea can be generalized to knowware development. Thus ordering instead of purchasing and buying is another mode of knowware acquisition. Actually, this is a form of knowware outsourcing in contrast of software outsourcing.

*How did we come to the idea of knowware?*

**KEYWORDS:**
KNOWLEDGE BASED SYSTEM, AUTOMATIC GENERATION OF KNOWLEDGE BASE, CONBES, BKDL, PSEUDO NATURAL LANGUAGE, PNL, TIANFENG, KONGZI, EBKDL, SELKAS, SELD, PROMIS, DODL, BIDL

Over the last two decades, we have spent considerable effort in developing various knowledge based systems. Our aim was to make the development of these systems automatic. After having developed several such systems, we reached to the conclusion that the development of a knowledge based system can be divided in two parts: the development of a relative stable system shell and that of the knowledge base. The key of automatic generation of a knowledge based system is the automatic generation of its knowledge base.

We started from automatic generation of expert systems. The first step was the construction of a platform called CONBES (CONstruct Book Expert System) for acquiring domain knowledge from technical books and generating expert systems automatically. Within the framework of CONBES, we developed a pseudo-natural language (PNL for short) called BKDL (Book Knowledge Description Language, for a definition of pseudo-natural language see below), which is a subset of Chinese. To transform a technical book in an expert system, we first copied the book text into computer and then modified it slightly to turn the natural language text to BKDL text. Each BKDL 'program' is compiled into a knowledge base with a hierarchy of frame representation. Combining this knowledge base with an existing expert system shell, we got a complete expert system. CONBES and BKDL have been used to produce expert systems automatically based on medical monographs[17].

The second experiment was the Tianfeng (heaven bees) system. It is used to produce ICAI systems automatically. The basic principle is also to acquire knowledge from textbooks and to generate a knowledge base automatically. However, Tianfeng differs from CONBES in the following points: 1. It produces teaching programs and tutorial texts rather than only doing consultations. 2. It produces exercises and is able to review and score user's answers. 3. It can revise a teaching program dynamically to fit individual users based on student modeling. There are two working versions: a Chinese version called KONGZI (Confucius) with a PNL called EBKDL (Education Book Knowledge Description Language), and an English version called SELKAS (Scientific English Literature Knowledge Acquisition System) with another PNL called SELD (Scientific English Literature Description). KONGZI and SELKAS share the same function module and the same knowledge base, which are language independent. Only their interfaces are different (Chinese resp. English)[18].

The third experiment was the PROMIS (PROtotyping MIS) system. This is a platform for acquiring business domain knowledge and generating MIS (Management Information System) like application systems automatically. The PNL used for this purpose is twofold: one for the use by domain experts is called DODL (Domain Ontology Description Language), another one for the use by MIS developers is called BIDL (Business Information Description Language), where "business information" means any information on the objective, organization, resource and activity that is relevant for the management of the business. In the terminology of PROMIS, business information is also called business situation. Domain experts write down domain knowledge in DODL texts, which will be compiled into domain knowledge base. After the knowledge base is established, any business user (e.g. CEO) can use BIDL to describe the general situation of his/her business. PROMIS translates the BIDL specification (which is a situation specification) into a requirement specification under support of the domain knowledge base. This requirement specification will be then further elaborated to generate an executable MIS, also under support of the same knowledge base. Note that this approach opens a third way of requirement engineering. As compared with the first way, which elicits requirement from end users, and the second way, which analyzes a written requirement specification, our third way generates a requirement automatically from a situation specification. This approach also means to replace the requirement analysis phase of the traditional waterfall model with a situation analysis phase. The key difference distinguishing PROMIS from other application examples listed above is the use of two different PNL: one for automatic generation of X knowledge base and another for acquiring business information from end users[21].

These experiments made us reflect on the following points:

1. The core of each application software is its domain knowledge.

2. If we ignore this core, then the remaining part is a relatively stable shell.

3. Each time when we need a new version of this application software, the best way is to replace the knowledge core with a new one and to keep the shell unchanged or modify it only slightly.

4. This means that the development of the shell can be separated from the development of the knowledge core.

5. This means knowledge core of application software may become a commodity independent of its shell.

These were the basic ideas that led us to the proposal of knowware.

*Can the production of knowware be made automatic?*

**KEYWORDS:**
ICAX, CAX SHELL, X KNOWLEDGE BASE, PSEUDO NATURAL LANGUAGE, PSEUDO NATURAL LANGUAGE UNDERSTANDING, KNOWWARE

The production of knowware can often be made automatic, because the production of knowledge base (or knowledge core in application software) can often be made automatic. This can be explained as follows:

Summarizing our above experiences in developing knowledge based systems, we get the following formula:

$$ICAX = CAX\ shell + X\ Knowledge\ base,$$

where any knowledge based system is represented as an ICAX system (Intelligent Computer-Aided X system), which is the integration of a relatively stable CAX shell with a X knowledge base, which can be changed easily. X means here anything, which can be done with help of computer. For example, an ICAD system consists of a CAD shell and a knowledge base with Design knowledge, while an ICASE system consists of a CASE (Computer Aided Software Engineering) shell with a Software Engineering knowledge base.

The introduction of X knowledge base has two meanings. The first meaning is to ensure the separability of the knowledge part from the operation part (calculation, reasoning, user models and user interface, etc.) in a knowledge based system. The second meaning is to ensure the interchangeability of the knowledge part.

Having the ICAX formula in mind, we can divide the development of knowledge based systems (ICAX systems) in three steps. First, we develop a CAX shell carefully, which will remain relatively stable in the ICAX life cycle. Second, we develop the X knowledge base by a process of prototyping, where the knowledge contained in X knowledge base will be acquired automatically or semi-automatically. Finally, we integrate the CAX shell and the X knowledge base together to get the ICAX system. The advantage of this approach is multiple. When the ICAX system needs to evolve, we may only change its X knowledge base, but not the CAX shell (or only a very little modification). We may work out domain standards for X knowledge base and buy it from other people whenever we need such, provided the standard is followed. We may also use some technique of automatic knowledge acquisition to produce the X knowledge base. The last point will be explained below.

Experts in knowledge engineering have been agreeing on the basic point of view that the most challenging problem in the construction of knowledge based systems is how to acquire enough and high quality domain knowledge. There are two basic sources of knowledge acquisition: knowledge contained in human brain and knowledge contained in some recording media, mainly in written documents. Usually people use knowledge elicitation techniques to acquire knowledge from domain experts and natural language understanding techniques to acquire knowledge from technical literature. As it was shown by practice, both approaches have disadvantages. Knowledge elicitation is often not successful because the experts are either not willing to cooperate or not able to summarize their knowledge. On the other hand, the scope and success rate of applying natural language understanding techniques is quite limited. A roughly 70% of correctness rate would be considered as normal. We decided to combine the both to develop a new methodology: the pseudo-natural language understanding approach of acquiring knowledge automatically. Below is a sketch of this approach:

1. Design a pseudo-natural language, which is similar to written natural language but can be parsed and understood by a computer (see the next question below);

2. Implement a compiler, which can parse pseudo-natural language texts, acquiring knowledge from it and organizing it in a domain knowledge base;

3. Each time when we want to translate a text book or a set of technical material in machine readable knowledge base, use an OCR device to scan the written document into the computer. Modify the scanned text slightly to turn it in its PNL form;

4. Let the computer parse and analyze the PNL text and produce a knowledge base (which may need to be

  integrated with an existing knowledge base). This is just the X knowledge base we want;

  5. Integrate the X knowledge base with the CAX shell to get the ICAX system.

Of course we can improve the above process to make it more perfect. We may present this 'rough' X knowledge based system to some domain expert who checks it with a set of test cases. With repeated knowledge refinement and improvement the experts can turn the 'rough' system into a refined one. We call this two staged process as a rational work division between human and computer. While the former reduces the difficulty of natural language understanding, the latter easies the formidable burden of extracting and summarizing knowledge from a huge amount of technical literature[17][18][19][20][21].

  Note that X Knowledge base can be considered as a preliminary form of KnowWare. On the other hand, knowware can be considered as the commercialized form of X knowledge base.

## *What are PNL and PNLU?*

**KEYWORDS:**

PNL, PNLU, PSEUDO NATURAL LANGUAGE, PSEUDO NATURAL LANGUAGE UNDERSTANDING, KEY STRUCTURE, SENTENCE PATTERN, CORE LAYER, DOMAIN LAYER, JARGON LAYER, ICAT, ICAIM, ICAMC

PNL is short for pseudo natural language and PNLU is short for PNL understanding. The former denotes a class of languages, while the latter denotes a kind of technique for processing PNL. Generally speaking, PNL looks very similar to natural language, but can be understood, analyzed and compiled by computer to an extent by which it can meet the need of some application, for example compiling a text book in a knowledge base for expert consultation.

1. Determine a set of semantic constructs of the application domain. For example, if the domain is mathematics, then the semantic constructs are case frames providing sentence semantics frequently used in mathematics textbooks, like concept definition, theorem proving, exercise presentation, etc.

2. Select a natural language, e.g. English, as background language;

3. Look for sentence patterns in this language, whose meaning corresponds to the semantic constructs selected in the first step. These sentence patterns may look like: **if \* then \* is called \***; **since \* is true and \* is not true we can infer from \* that \* is true**. A sentence pattern is also called a keyword expression, where a keyword is a contingent symbol string without stars in a sentence pattern.

4. The correspondence between sentence patterns and semantic constructs may be many to one. For each semantic construct, select only one sentence pattern as its syntactic representative and remove all other sentence patterns (with the same meaning) from the language. Call the selected sentence pattern the normal form among all sentence patterns having the same meaning. Call the set of selected sentence patterns and their combination (see below) the key structure of the language, which is now called pseudo-natural.

5. Each sentence pattern of the key structure has a specified syntax, semantics and pragmatics. That means

its meaning is well determined. The semantics of all other contents (the parameters of the sentence patterns, i.e. those parts marked with stars in 3. above) is only interpreted in context of the sentence patter they are reside in. For example, the semantics of the sentence pattern **if * then * is called *** can be explained as follows: the first star is the condition, the second star the father concept and the third star the concept itself. Its pragmatics is concept definition and classification.

6. The key structure is defined in form of a grammar. That means: not every combination of sentence patterns is a legal key structure.

7. Use a domain knowledge base to specify the syntax, semantics and pragmatics of each sentence pattern in the key structure in order to support computer understanding of pseudo-natural language texts.

While parsing a PNL text, the computer tries to understand the text, based (and only based) on the semantics of the underlying key structure. This understanding is necessary superficial in the sense that no information other than that implied by the key structure will be gained by the computer. It abstracts away all unnecessary details regarding the current application and thus makes the language understanding much easier. For example, consider the sentence:

**If** the color of the blood cell is red **than** the blood cell **is called** erythrocyte.

This is a (shallow) definition of red blood cell. But here we can already see the abstraction principle of PNL. With this knowledge, a computer can answer questions like "what is erythrocyte?" , "How do we call a blood cell when the color of the blood cell is red?" etc. even without knowing the meaning of "red" or "cell".

It is easy to see that if we enlarge the key structure of PNL, then the computer will acquire more detailed knowledge from a PNL text. For example, if we add the sentence pattern "**color of \* is \***" to the key structure, then the computer may additionally know that color is an attribute, which can be used to describe physical objects (by defining its pragmatics in the knowledge base). On the other hand, if we reduce the key structure, then the knowledge acquired by the computer will have a larger granule. In this way, PNL defined with different key structure form a spectrum, which is a partial order. The upper limit of this spectrum is the natural language, while the lower limit is the formal language consisting of meaningless strings of symbols.

Since each PNL is domain oriented, one would ask the question: should we design a particular PNL for each domain? If the answer were yes, then the work of defining new PNL would be a heavy burden. Fortunately we can divide the key structure of a PNL in several layers. There are three basic layers: the core layer, which contains sentence patterns used in all domains; the domain layer, which contains technical expressions used in a particular domain; and the jargon layer, which contains professional expressions used by a particular group of users. Each time when we design a new PNL, the core layer, which occupies the major part of the key structure, does not have to be modified. Only part of the domain layer should be renewed. The jargon layer is usually very small and does not play an important role. Of course it is also possible to define intermediate layers between the basic layers.

The development of ICAX systems mentioned above is all based on PNLU techniques. The expert systems generated by CONBES platform are ICAMC (Intelligent Computer Aided Medical Consultation) systems ($X = MC$). The PNL used is BKDL. Each BKDL 'program' is structured in three levels: sentence pattern, sentence type (sentence patterns of similar pragmatics, e.g. the classification type) and sentence chapter (knowledge chunk consisting of sentence patterns of different types)[17]. The Tianfeng platform generates ICAT systems (X = T). The corresponding PNL are EBKDL (acquiring knowledge from Chinese texts and generating Chinese version ICAT) and SELD (do the same thing in English). The knowledge representation is frame based[18]. The PROMIS platform generates ICAIM systems (X = IM = Information Management). The corresponding PNL are DODL (for domain experts) and BIDL (for end users). The knowledge is ontology based[20][21].

In summary, it is much easier to adapt written scientific materials into PNL texts than to elicit the knowledge from domain experts, and once this is done, then tasks in sequel, such as text understanding, X knowledge acquisition, X knowledge base construction, ICAX system generation, etc, can be accomplished automatically.

*What is the difference between knowware, knowledge base, X knowledge base and expert system?*

**KEYWORDS:**
KNOWWARE, KNOWLEDGE BASE, X KNOWLEDGE BASE, EXPERT SYSTEM

First we have to point out that among these four candidates only expert system is a kind of software, all other are only data packages, not software, because knowware and X knowledge base don't have a program controlling the use of the knowledge, and a knowledge base is always a part of a software containing it. On the other hand, besides knowware itself, all other three candidates are not knowware.

Generally, a knowledge base and its management system (in case of expert system: a knowledge base and its inference engine) are like a married couple such that you can not keep one of them unchanged while replacing the other by a new one. But this is possible with an X knowledge base, where the interchangeability is one-sided. One can change the X knowledge base while keeping the CAX shell unchanged, but not the other way round. Knowware is more flexible than X knowledge base such that it can serve as plug-in spare parts for any software that has standard interface for knowware embedding.

Table 2 shows the difference between these four concepts.

|  | KNOWWARE | X KNOWLEDGE BASE | KNOWLEDGE BASE | EXPERT SYSTEM |
|---|---|---|---|---|
| Is-a software | No | No | No | Yes |
| Has control mechanism | No | No | Partially (KB may have management system) | Yes (ES always has inference engine) |
| Standard interface | Yes | Partially (may have local standard) | No | No |
| May work as plug-ins | Yes | Partially (for fixed CAX shells only) | No | No |
| commercialized | Yes | No | No | Yes |

Table 2. Difference between Knowware, X Knowledge base, Knowware base and Expert System

*Why don't we use the term "intelligent ware"?*

**KEYWORDS:**
INTELLIGENT WARE, RODNEY A. BROOKS, MARVIN MINSKY

Knowware is directly related to intelligence: if software is considered as condensation and crystallization of human knowledge, as pointed out by late prof. Xiwen Ma, then knowledge could be considered as condensation and crystallization of human intelligence.

Because of the reason said above, one might think that we are equally right to use the name intelligent ware (instead of knowware) to denote the knowledge part separated from software. But there is a remarkable difference between intelligence and knowledge. While knowledge can be separated into different parts, intelligence is inseparable according to the opinion of AI experts. In his IJCAI '91 award paper "Intelligence without reason", Rodney A. Brooks proposed four principles of behavioral intelligence. When talking about the fourth principle, the emergence principle, he cited Marvin Minsky's argument that "there is never any heart in a program" and concluded that within a computer program "it is hard to point at a single component as the seat of intelligence"[4]. Therefore, an intelligent program is intelligent as a whole. It is not reasonable to separate a part from a program and call it intelligent.

*What kinds of architecture does a knowware have?*

**KEYWORDS:**
ARCHITECTURE OF KNOWWARE, MONOLITHIC ARCHITECTURE, DATA LIST ARCHITECTURE, DOLPHIN, INFERENCE ARCHITECTURE, HETEROGENEOUS ARCHITECTURE, ONTOLOGY, CBS AGENT, RELATIONAL TABLE, MIXWARE

Depending on the problem to be solved and the knowledge contained in it, the architecture of knowware can take various forms.

Generally speaking, it is different to talk about the architecture of knowware or architecture of knowledge organization contained in knowware. Nevertheless, at this place, we mean always architecture of knowledge organization when we talk about architecture of knowware.

The first and simplest form of knowware architecture is the monolithic form. In this form, knowledge is considered as a whole and will not be decomposed in blocks or pieces. The first of our examples given above, a digitally coded song in a MP3 player, takes this form.

The second form of knowware architecture is a data list. Here we mean a data list in generalized sense. This data list can be heterogeneous. It may be a data table, or a relational data base, or a data warehouse. A data list can be used with a knowledge middleware, which is a search or analysis tool. In the Dolphin system, the information of 0.76 Million papers is stored as relational tables about author, title, abstract, keywords, text, reference, publication time, publication place, etc.[38]

The third form of knowware architecture is inference architecture, like that used in a traditional expert system. Three of the most widely used inference architectures are production system, logic program and artificial neural network. Inference architecture is used whenever the user problem can only be solved through a reasoning process. This form of knowledge architecture is particularly suitable for representing problem solving tasks, like business intelligence of ERP mentioned above.

The fourth form of knowware architecture is the combination of a data list with inference architecture. Knowledge structures like objects, agents and ontologies belong to this category. Here we mention the Pangu structure of commonsense knowledge, which consists of three types of knowledge representation: CBS agents, ontologies and relational tables[22]. Both agents and ontologies form two dimensional data lists (trees) hierarchically organized with inheritance mechanism. Ontology is a directed graph of agents, where each edge is marked with a semantic relation. The third type of knowledge representation, the relational tables, is used to support a high efficiency. Agents and ontologies in form of relational tables are at least 10 times more efficient in inference than those in original data list form.

The fifth form of knowware architecture is a heterogeneous structure consisting of many substructures of different kinds. A typical example is the courseware containing knowledge modules of different types implementing different functions. For example, a courseware for teaching mathematics may contain mathematical texts, mathematical formula with exotic symbols, exercises with answers, student models, proof techniques, etc. Each of these knowledge modules may have different architecture. Therefore it may be not suitable to include all these in a single heterogeneous knowware architecture. It would be much better to implement these knowledge modules with different knowware and then combine all the knowware together to get a nested knowware component. We will discuss this problem below in the section about mixware.

*Which knowledge schema is used for knowware?*

**KEYWORDS:**
KNOWLEDGE SCHEMA, KNOWLEDGE SOURCE, KNOWLEDGE ORE, KNOWLEDGE MAGMA, KNOWLEDGE CRYSTAL, KNOWWARE, KNOWLEDGE MIDDLEWARE, WATERMARKING

Knowledge schema is different from knowledge representation. Knowledge representation is concrete, like production systems, logic programs, semantic networks, neural networks, etc., while knowledge schema is abstract. It does not relate to any concrete representation form, but is focused on general knowledge organization principles. These principles include: granule size of knowledge elements, modularity of knowledge clustering, hierarchy of knowledge leveling, interconnection of knowledge reasoning, etc. We differentiate knowledge schema in two development stages. The first one involves knowledge schema for storing, archiving and organizing knowledge, used in half fabricates of knowware. The second stage involves knowledge schema used in knowware itself.

Knowledge schema of the first stage is divided in four layers: the knowledge source layer, whose schema is in any unstructured form, of which the examples are the immense number of Web pages; the condensed knowledge source (also called knowledge ore) layer, whose schema is semi-structured and organized according to its knowledge content, of which the examples are digitalized books, technical reports, encyclopedia, etc.; the knowledge magma layer, whose schema is organized in encyclopedia principles with small granularity and the knowledge crystal layer, whose schema is structured according to the habitudes of the domain experts, for example the tables of genes found from certain animal's DNA sequence. As for knowledge schema of the second stage (knowware), it is often heterogeneous and crossing several domains, possibly with some additional features, such as watermarking to protect the knowledge patent and to guarantee that the knowledge is verified by some authority to be correct and up to date.

However, the statement "schema is structured according to the habitudes of the domain experts" is quite rough and even fuzzy. In fact, this statement means that the schema of the knowledge contained in knowware is not limited to any particular form. Here the key is the knowledge middleware, which we will explain below. One of the functions of knowledge middleware is to translate one representation into another. It is like a voltage adapter or a file format translator. Due to this reason, one does not care the concrete representation used by a particular knowware, if one has suitable knowledge middleware to do the translation. Nevertheless, knowware developers are not encouraged to use arbitrary knowledge representation at their free will. They'd better to select one of the standard representations accepted by domain experts. Otherwise they'd need to develop the corresponding knowledge middleware themselves.

## *What is knowledge middleware?*

**KEYWORDS:**
KNOWLEDGE MIDDLEWARE, KNOWLEDGE BROKER NETWORK, KBN, KNOWLEDGE EXTRACTION MIDDLEWARE, KNOWLEDGE TRANSFORMATION MIDDLEWARE, KNOWLEDGE CRYSTALLIZATION MIDDLEWARE, KNOWWARE PRODUCTION MIDDLEWARE, KNOWWARE OPERATING MIDDLEWARE, KNOWWARE COMBINATION MIDDLEWARE, KNOWLEDGE SERVICE MIDDLEWARE, KNOWLEDGE PUMP, KNOWLEDGE DRESSING, PNLU, WKPL, KNOWLEDGE KIDNEY, KNOWLEDGE CRYSTAL, OCCAM'S SHAVER, KNOWLEDGE DECOMPILATION, KNOWLEDGE SERVER, KNOWLEDGE GRID, NKI, SEMANTIC WEB, KNOWWARE BASED WEB SERVICE

The category of software can be classified in system software and application software. For knowware there is no such classification. Every piece of knowware is application oriented. There is no system knowware. Nevertheless we will introduce a class of important software which plays the role of "system knowware": the knowledge middleware. Thus "system knowware" is software. Knowledge middleware is a class of software tools accompanying the whole lifecycle of knowware. The development, application and management of knowware involve knowledge acquisition, selection, fusion, maintenance, renewing and many other functions. Knowledge middleware is used to perform these jobs. It is different from the conventional middleware concept in software engineering. Software middleware helps application programs to work cooperatively in a networked heterogeneous environment. The operation of knowware needs a network in a broader sense. This functional network connects not only knowware with knowware, but also knowware with software, knowware with knowledge source and knowware with human users. We call it the knowledge broker network, KBN for short. Thus, knowledge middleware is the underlying set of software tools based on KBN and a set of knowledge protocols, whose function is to support the effective development, application and management of knowware.

Roughly classified, we have the following classes of knowledge middleware (KM for short): knowledge extraction KM, knowledge transformation KM, knowledge crystallization KM, knowware production KM, knowware operating KM, knowware combination KM and knowledge service KM. Table 3 lists some main functions and techniques of knowledge middleware.

| | **FUNCTIONS** | **TYPICAL TECHNIQUES** |
|---|---|---|
| Knowledge Extraction Middleware | knowledge extraction from knowledge source | knowledge pump, knowledge dressing, PNLU, WKPL, text and data mining, deep search, machine learning |
| Knowledge Transformation Middleware | transformation between different knowledge formats and representations | compiler, translator, transducer |
| Knowledge Crystallization Middleware | selection, updating, fusion, evolution and organization of extracted knowledge elements | ontology structuring and alignment, knowledge management, knowledge kidney, contradiction resolution, truth maintenance, noisy information removal, fuzziness clarifying |
| Knowware Production Middleware | cut knowledge crystals using an Occam's shaver and integrate them according to requirement | requirement analysis, component engineering, rapid prototyping |
| Knowware Operating Middleware | knowledge analysis, synthesis, knowledge view generation, knowledge publication, user friendly knowledge representation | statistics, diagram techniques, text mining, content summary and report generation |
| Knowware Combination Middleware | acquire a set of basic knowware based on user requirement and assemble them into more complicated knowware | knowledge modeling, automated architecture design, component engineering, mixware engineering |
| Knowledge service Middleware | register, enroll, distribute, and protect intellectual property, and fee collection; instant and long term user online problem solving | requirement analysis, Web service, knowledge service, e-commerce, knowware based Web-service, semantic Web, knowledge grid[39], NKI[5] |

Table 3. Knowledge Middleware Classification

*What is knowware engineering?*

**KEYWORDS:**

KNOWWARE ENGINEERING, KNOWLEDGE MIDDLEWARE, COMPUTER AIDED KNOWWARE ENGINEERING, CAKE, LIFE CYCLES OF KNOWWARE ENGINEERING

We may have two definitions of knowware engineering that are not identical. The more scientific definition is: knowware engineering is the systematic application of knowledge middleware with the goal of knowware generation, evolution and application.

Another definition, which is more engineering oriented, is as follows: If computer can generate and manage the use of normalized, packaged, and commercialized knowledge, i.e, knowware, this process of knowware generation and management is called knowware engineering. Obviously, a more suitable name for the second definition is computer aided knowware engineering, CAKE for short.

Knowware engineering has life cycles, just as software engineering does. Depending on how one obtains knowledge, organizes it in knowledge crystals, maintains it, makes it evolving and transforms it to knowware, one has different kinds of life cycles of knowware engineering.

*What are the life cycle models of knowware engineering?*

**KEYWORDS:**

LIFE CYCLE MODEL, FURNACE MODEL, CRYSTALLIZATION MODEL, SPIRAL MODEL, KNOWLEDGE ORE, KNOWLEDGE MAGMA, KNOWLEDGE CRYSTAL, KNOWLEDGE SEA, KNOWLEDGE PUMP, KNOWLEDGE KIDNEY, KNOWLEDGE SPIRAL, TACIT KNOWLEDGE, EXPLICIT EXPERIENCE, KNOWLEDGE DISCOVERY, DECISION MAKING

Scholars have suggested many development models in software engineering research, which are closely related to software life cycle, such as the well-known waterfall model, fountain model, spiral model, and rapid prototype development model, etc. Knowware engineering should have its own development model and life cycle.

The first model of knowware life cycle is the smelting furnace model, furnace model for short. A smelting furnace accepts and smelts raw material inputted in a batch way, like the blast furnace smelts iron ore, or the steel furnace smelts iron blocks. In the knowware practice, this smelting furnace is a complicated knowledge processing system with a massive and heterogeneous knowledge base containing knowledge magma. Each time when knowledge ore (books, leaflets, newspapers, technical reports, etc.) is inputted, it will undergo a process of ore dressing, where knowledge ore will be decomposed in smaller knowledge elements.

These elements, though organized in some way, form the knowledge magma of the furnace. Each time when there is a knowware requirement, the needed knowledge elements will be extracted from the knowledge magma to form the knowledge crystal. The knowledge crystal will be grown in a repeated process of knowledge extraction, integration and reorganization, until it is ready to be commercialized to become a product. Figure 3 is an illustration of the furnace model.

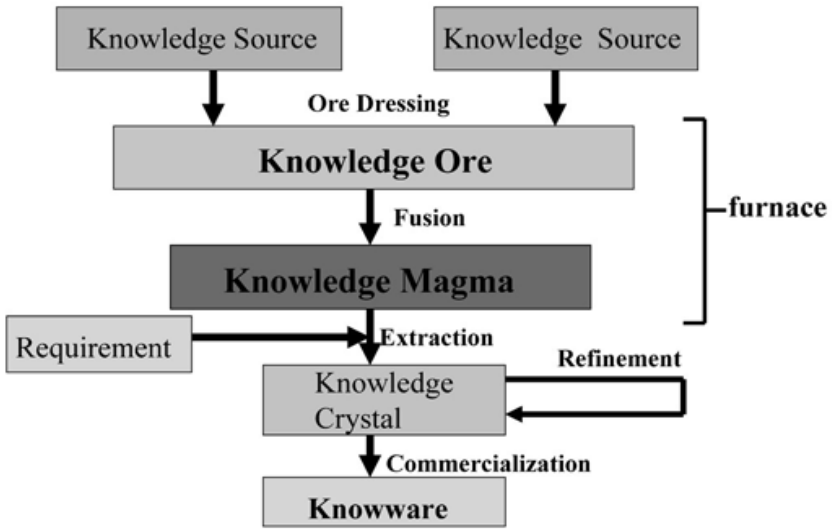

Figure 3. The Furnace Model

To give an example, we recall the project of producing ICAI systems mentioned above. It automatically acquires knowledge from a set of imported textbooks and technical leaflets. These books and leaflets function as "knowledge ore", which will be broken in small knowledge units and smelted to knowledge magma in the knowledge base. That is, their knowledge will be extracted and reorganized in a ready to use form. Each time a new ICAI is requested by some individual, the relevant knowledge will be selected and reorganized in a knowledge crystal – the teaching course. Just as in the real smelting process, different "impurities" have to be added or removed to get quality products, the same thing will happen in this knowledge crystal production process[18].

The second model of knowware life cycle is the crystallization model. Its basic idea can be described as follows. Knowledge required by certain knowware is often distributed and scattered in a large information space. The knowledge acquisition process has to go through a process of acquisi-tion, refinement, analysis, fusion, and reorganization in order to be applicable. In most cases, this process of knowledge gathering does not have a absolutely perfect end. For any topic to be studied or researched new and useful knowledge is accumulated continuously, while old and obsolete knowledge is discarded constantly. Imagine you are entering the National Library. You may want to find out all people who have made contribution to the construction of the city you are now living in. You look for each trace of such contribution bit by bit and try to sort the set of information you have obtained to figure out a global picture. We call the immense information space (in the above example the National Library) a knowledge sea, and the knowledge formation process a process of knowledge crystallization. The demand and making of knowledge is like a center of crystallization. The knowledge in the sea precipitates and gathers around this center, so that "crystals" become bigger and bigger. The structure of knowledge crystals is the rules and norms of knowledge representation and organization. Another typical knowledge sea is the knowledge hidden in the World Wide Web [13][38].

In the process of crystallization, the new and original knowledge may not be always compatible or in harmony. The knowledge in crystals might be degraded and obsolete, new knowledge is needed to replace or modify the original knowledge. Contradictions in knowledge will be resolved in favor of more dependable and reliable knowledge. Fuzzy and inexact knowledge will be clarified and refined. Finally, inappropriately organized knowledge will be reorganized. To knowledge crystals, this is a process of annealing and recrystallization. Therefore, knowledge crystals are dynamic and metabolic objects. On the other hand, existing knowledge crystals can enter the knowledge sea as new knowledge particles, so that they can be used again in a new and higher level.

In summary, we need a control mechanism, called knowledge pump, to implement the process of knowledge gathering. A knowledge pump has to accomplish the tasks of knowledge extraction and knowledge filtering. We need another control mechanism, called knowledge kidney, to implement the process of knowledge maintenance and knowledge renewing. Figure 4 presents the crystallization model of knowware development.

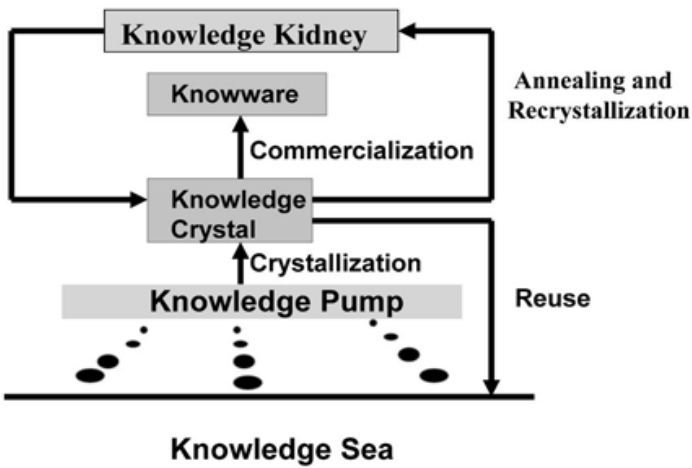

Figure 4. The Crystallization Model

The third model of knowware life cycle is the spiral model, which characterizes the formation and transformation of experience (tacit knowledge) and knowledge (explicit experience), as well as knowledge discovery and decision making.

It circles the loop:

business rule $\xrightarrow{business-practice}$ business data $\xrightarrow{data-mining}$ informative message

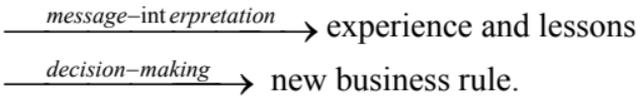 experience and lessons
$\xrightarrow{decision-making}$ new business rule.

In the knowware practice, this knowledge spiral describes the spiral evolution of expert knowledge (from experience to theory). That means the knowledge spiral can serve as a model of knowledge crystal formation. Compared with the crystallization model, the knowledge spiral model puts more focus on improving the knowledge quality than on increasing the knowledge amount. This model of knowware development is particularly suitable for developing private knowware, where the user of knowware is often involved in the knowledge accumulation and improvement process. Figure 5 shows the spiral model.

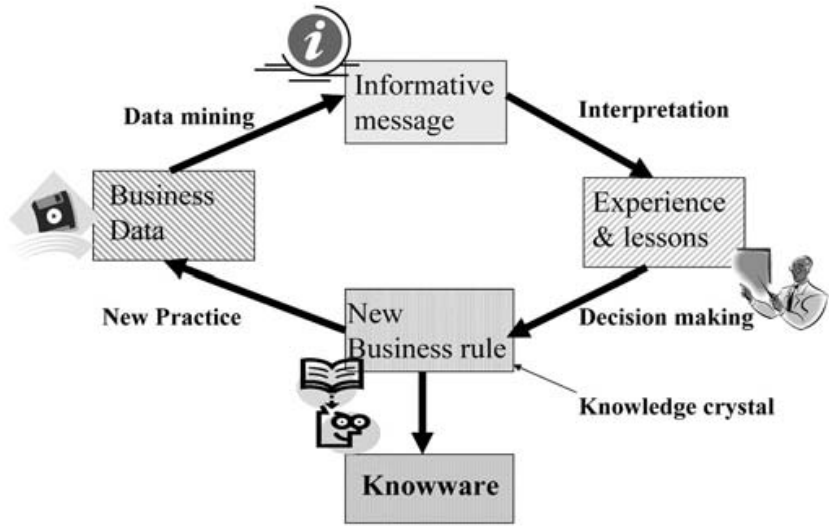

Figure 5. The spiral model

*What is the form of a knowledge pump and how does it work?*

**KEYWORDS:**
KNOWLEDGE PUMP, KNOWLEDGE SEA, XML PLUS PNLU TECHNIQUE, WORLD WIDE WEB, MARKUP LANGUAGE, DOMAIN ORIENTED MARKUP LANGUAGE, KEY STRUCTURE, WKPL

A knowledge pump collects knowledge elements from a usually immense information space (knowledge sea). There are many well known techniques of knowledge acquisition, including machine learning, inductive programming, data analysis, knowledge elicitation, etc. Each of these techniques can be used to build knowledge pumps for practical use. In this book there is no space to list all of them in detail. We want only to sketch the idea of a new methodology for constructing knowledge pumps based on a XML plus PNLU technique.

Consider the World Wide Web. Google Inc. has announced in year 2004 that they have indexed 6 billion items, including 4.28 billion Web pages, 880 million images, 845 million Usenet images, and a large set of other information sources[13]. We need powerful tools to implement this knowledge pump, which should be able to distinguish wanted knowledge elements from all unwanted data and information of this immense knowledge sea. A knowledge pump is something like fishing net. The kind of fishes caught by fishing net depends on the size of meshes. The smaller the size of meshes is, the smaller is the size of fishes caught by the net. Here the size of fishes means the granule of knowledge acquired by the knowledge pump.

As a first choice, the XML based document markup technique can be used to implement knowledge pumps. XML based document representation provides a possibility of analyzing the concept structure of a Web page or other files. RDF and OWL have the additional advantage that certain semantics can be assigned to the structured elements. In this sense, a markup language may have a new function of knowledge mining. The difference is: as a markup language it is attached to knowledge source, while as a knowledge mining language it is attached to browser. With help of XML devices like tags and RDF, OWL definitions, the knowledge pump can locate the knowledge elements needed for the current task. We take XML as our first layer of tools for knowledge pumping.

However, the XML framework in its general sense is not necessary the best candidate to be used by a knowledge pump. Its tags definition depends on individual XML document developers. Different people may use different tags to denote the same thing. This makes the knowledge pump having difficulty to identify wanted knowledge elements. Some standard is needed. Fortunately there are already some domain oriented markup languages based on XML, e.g. in the field of mathematics, chemistry, medicine, biology, etc. Thus, the second layer of knowledge pumping tool is the set of all domain oriented markup languages.

Both techniques mentioned above can only locate a subset of the immense information space, but cannot fetch out the knowledge elements contained in it. We decided to combine our PNLU technique with XML based document markup languages to implement a third layer of knowledge pumps. Different to the XML based markup languages, the PNLU technique does not invent new tags to add to the marked text. It just presumes a key structure of PNLU for the tagged space to declare that 1. Each piece of the tagged space mined by the knowledge pump should match the specified key structure at least partly. 2. All contents of a tagged space, which do not meet any part of the key structure, will be discarded. 3. Those contents matching the key structure will be decomposed in basic knowledge elements according to key structure semantics and archived in the knowledge base.

Now we give an example to illustrate the use of this technique. Assume we want to search knowledge about the concept "software". We may use the XML based knowledge pump to perform this job:

<SOFTWARE>
    \*\*\*
</SOFTWARE>

where the knowledge to be found is marked with three stars. The knowledge pump will look for all XML texts tagged by <SOFTWARE> and </SOFTWARE>. However, what one gets by using this knowledge pump is only a piece of text like the following:

<SOFTWARE>
Software is a set of programs running on computer with corresponding documentation. Software is classified in three classes: system software, application software and supporting software. System software includes operating systems, compilers, database management systems and utility programs. Application software includes software for numerical computation, expert systems, etc. Supporting software includes software middleware, application server, etc.
</SOFTWARE>

which is not computer understandable without natural language understanding.

However, by using the technique of PNLU, we can write the knowledge pump with the key structure inserted, like the following:

<SOFTWARE>
    \*\*\* is a \*\*\*.
    \*\*\* is classified in \*\*\* and \*\*\*.
    \*\*\* includes \*\*\* and \*\*\*.

```
         *** includes *** etc.
   </SOFTWARE>
```
This knowledge pump will not only locate the text containing knowledge about software, but also fetch out knowledge elements characterized by this key structure. The result will be

```
   <SOFTWARE>
```
Software **is a** set of programs running on computer with corresponding documentation. Software **is classified in** three classes: system software, application software **and** supporting software. System software **includes** operating systems, compilers, database management systems **and** utility programs. Application software **includes** software for numerical computation, expert systems, **etc.** Supporting software **includes** software middleware, application server, **etc.**

```
   </SOFTWARE>
```
where the boldface characters denote the key structure.

This knowledge pump returns the following information: 1. a text relating to the definition of the concept "software"; 2. a set of knowledge elements obtained by decomposing this text; 3. a semantic hierarchy of knowledge obtained, which looks like follows:

1) An (intensional) definition of software (software **is a** …..);

2) A classification of software (software **is classified in** ……);

3) An (extensional) definition of system software, application software and supporting software (system software **includes**…….).

This is only a very simple example. More complicated application is possible. It is to note that to certain degree, PNL also embody the format of knowledge crystallization. By inserting key structure in XML text we can not only control the content of knowledge extraction, but also its granularity.

We call our markup language WKPL (Web Knowledge Pumping Language), which is used to specify knowledge requirement and acquire knowledge over the Web. WKPL is a generalized markup language. A XML like markup language uses prefix tags like <SOFTWARE> and postfix tags like </SOFTWARE> to markup a text. WKPL uses in addition key structure based infix tags like "**is a**" to markup a text. WKPL is actually a family of languages. The key structure implied in a WKPL program specifies what kind of knowledge to be acquired and how to compile and edit the acquired knowledge. By running a WKPL program, one can obtain a knowledge crystal from the Web that meets the specific requirement.

*What is the difference between knowware and knowledge crystal?*

**KEYWORDS:**
KNOWWARE, KNOWLEDGE CRYSTAL, KNOWLEDGE HOMOGENEOUS, LENAT, MICRO-THEORY, INTELLECTUAL PROPERTY

A knowledge crystal can be understood as a half fabricated and knowledge homogeneous form of knowware, which is a well recognized and organized set of knowledge restricted to some domain. Here the word 'domain' is generally understood as a scientific discipline. Knowledge homogeneous means the knowledge contained in the knowledge crystal is limited to this scientific discipline, for example organic chemistry. It can be considered also as a consistent but not necessary complete knowledge module (compare with the micro-theory in the terminology of Lenat[16]) of a formatted and modularized knowledge base. The latter may be neither consistent nor complete. Both knowledge crystals and the knowledge base are subject to a steady evolution. A knowledge crystal is usually not a commodity and does not have to take care of the commercial standard. Besides, it is more general-purpose oriented than a knowware, which has a relatively narrow focus of application. It is like the half-fabricated fish and pork in the kitchen of a big restaurant, while a knowware is a well prepared dish made of a set of different half products. As half-products, knowledge crystals require further processing to become knowware. Table 4 summaries the major differences between knowledge crystals and knowware.

|  | **KNOWLEDGE CRYSTALS** | **KNOWWARE** |
|---|---|---|
| Integrity Principle | The interconnection of knowledge within a specific domain | The usefulness of knowledge under a specific requirement |
| Classification Principle | Knowledge of different domains usually formulate different knowledge crystals | Knowledge of different domains can be combined into a unified knowware |
| Organization Principle | Emphasize scientific rules | Emphasize efficiency in applications |
| Production Principle | Driven by development and evolution of knowledge | Driven by user and market requirements |
| Property Right Principle | May not have intellectual property | Always have intellectual property |
| Update Principle | Updated partially as soon as the knowledge is renewed | Updated in form of replacement by a new knowware (, which is generated by an evolved knowledge crystal) |
| Update Cycle | Continuous evolution | Version-wise evolution. Knowware will be renewed only after an extended accumulation and evolution of knowledge crystal |

Table 4. Differences between Knowledge Crystals and Knowware

## *What is knowledge view and knowledge interface?*

**KEYWORDS:**
KNOWLEDGE VIEW, KNOWLEDGE INTERFACE, DATA VIEW, KNOWLEDGE PROTOCOL, KNOWLEDGE ADAPTER

The same knowledge can be observed from different aspects. This function of knowledge processing is called knowledge view. It is generated by knowledge middleware according to user needs. Compare it with the data view concept in database technology. A data view selects only those data items from a huge data base, which are of interest for the current user, and at the same time masks all other data. For example, imagine that all information of the professors in a university is stored in a relational database. A reader of the educational office is only interested in "give lectures" attribute of each professor, while a reader of the financial office may be only interested in the "salary" attribute of the professors. The data view filters out all uninteresting attributes and provides only those attributes the current reader is interested in. A knowledge view may be more complicated than the data views, since usually the knowledge representation is more complicated than a relational database.

Knowledge interface is different from knowledge view. Each knowledge interface is attached to some knowware. It specifies a set of knowledge protocols, telling the content of knowware and how to use it. It includes at least a list of content, format of knowledge representation, possible applications of the knowware, version number, watermarking and authentication provided by validation center, list of knowledge middleware applicable to it. A high level knowledge interface may contain a knowledge adapter for transforming knowledge representation when it is installed.

*How to combine software development and knowware development together?*

**KEYWORDS:**
PARADIGM, ICAX, CAX SHELL, X KNOWWARE, X KNOWLEDGE MIDDLEWARE, TAX MANAGEMENT SYSTEM, DOLPHIN, X MIXWARE, ERP

There are three different paradigms of combining software development with knowware development: fixed software and changeable knowware; fixed knowware and changeable software; cooperative development of software and knowware. We still use X to denote some application domain. The first paradigm can be represented by the formula ICAX = CAX shell + (X knowware), which is the advanced form of the old formula ICAX = CAX shell + X knowledge base. In this paradigm, the CAX shell is kept fixed, while the X knowware can be changed whenever the X knowledge is renewed. For example, the tax law knowware in a tax management system must be changed when the government publishes new tax rules.

The second paradigm can be represented by the formula ICAX = X knowware + (X knowledge middleware). In this paradigm, the X knowware is kept fixed, while the X knowledge middleware can be changed whenever one needs a new function of the same X knowware, since new function needs new knowledge middleware. In order to be convinced about that, just have a look at the Dolphin system introduced above.

The last paradigm can be represented by the formula ICAX = X knowware + X knowledge middleware + software. It develops knowware together with knowledge middleware and software. A complex consisting of all these three kinds of modules is called mixware. Therefore we could also replace the left side of the above formula with X mixware. In this case, the whole X mixware is developed by a cooperative working team under a unified planning. For example, the business intelligence of an enterprise will be transformed into a knowware that will form a part of the ERP of that enterprise. But ERP is a very complicated system. We call this paradigm the paradigm of software/knowware co-engineering or simply mixware engineering.

*What is the model of software/knowware co-engineering?*

**KEYWORDS:**
SOFTWARE/KNOWWARE CO-ENGINEERING, MIXWARE, GLUING MIDDLEWARE, OPERATIONAL SOFTWARE, DEVELOPMENT THREAD, PARALLEL THREADS, LIFE CYCLE PHASE, PHASE BUS, REQUIREMENT BUS, DESIGN BUS, IMPLEMENTATION BUS, WATERFALL MODEL, LADDER MODEL

The co-engineering process applies to the development of complicated mixware. It differs from the traditional software engineering process in many aspects. First, it is a mixed process involving both software engineering and knowware engineering issues. In addition, the knowledge middleware issues are also considered as a bridge connecting the two sides. Second, the global system requirement will be split into three partial requirements, which initiate three parallel threads (knowware, gluing middleware, pure operational software) of system development. Third, each development thread is divided in its own life cycle phases. Fourth, the corresponding life cycle phases of all development threads are connected by phase buses to enable information exchange (communication) between development threads. For example, there is a requirement bus connecting the requirement phases of all development threads. Fifth, the development threads and the phase buses form a net with the former as meridians and the later as parallels. Each cross point of this net is a check point of the engineering. They are used to assure the integrity and consistency of products and half products on the confluent places of the three parallel development threads. Sixth, feedback information during system development is not only passed along development threads in the vertical direction, but also passed along phase buses in the horizontal direction (for example, the source of a bug found in software design may hide in knowware design and vice versa). As a result, we have the following software/knowware co-engineeing process:

1. Requirement specification of the whole system,

2. Requirement decomposition along requirement bus (in software requirement, knowware requirement and gluing middleware requirement),

3. Communicating analysis of the above three requirements (analysis of the three requirements is synchronized by communication along requirement bus),

4. Synchronous development of requirement bus into design bus (by transforming three requirements in three designs concurrently),

5. Communicating integration of three designs,

6. Synchronous development of design bus into implementation bus (by coding the three designs concurrently),

7. Synchronous test and verification of three implementations,

8. Integration test and validation of the whole system.

This model is a variation of the waterfall model of software engineering in the sense that it also undergoes a stepwise refinement from the requirement analysis downwards until system generation. But they are different in the number of development threads, in the use of phase buses and in the directions of information feedback. One may think that we can use a model of three parallel waterfalls. But a multiple waterfalls model is not appropriate, either, because there is no connection between the waterfalls and thus no communication between the parallel threads. Therefore we use a ladder model to characterize this process. This model binds two ladders together to form a combined ladder with three handrails, like a wide staircase with one handrail in the middle and two on both sides. While developing a mixware, three development teams climb down the three handrails concurrently. Whenever any team finds a bug, this team may climb its handrail up, or send a message towards left or right to the other two handrails to let the other teams climb up. This action may take place for an unlimited number of times, until the source of the bug is found. Generally, they may go a zigzag line to find the problem rather than just a straight line in traditional waterfall model. This is why we call it a ladder model. An illustration of the ladder model is shown in Figure 6.

Certainly this is not a brand new life cycle definition of information system engineering. One can find quite a few impacts from the software engineering concepts and techniques in the above paraphrase. However, there are special difficulties raised by this co-process definition, which do not occur in traditional software engineering process techniques.

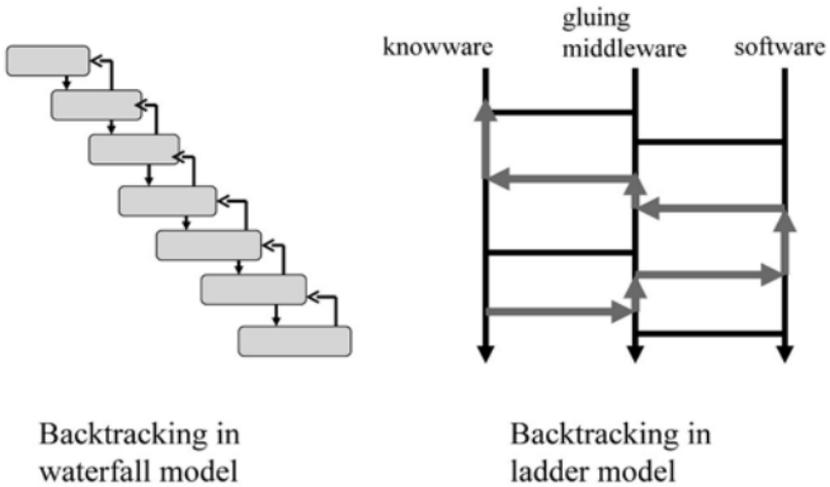

Figure 6. Compare ladder model with waterfall model

*What is mixware?*

**KEYWORDS:**
MIXWARE, COORDINATOR, GLUING MIDDLEWARE, KNOWLEDGE MIDDLEWARE, SOFTWARE CONNECTOR

Roughly speaking, mixware is an organized complex combining software and knowware. Coordinators, called gluing middleware, are used to connect software and knowware components in a mixware. The coordinators connecting a knowware component with other components are called knowledge middleware. Those connecting software components only are called software connectors.

*How to adapt object-oriented approach in software engineering to object-oriented approach in mixware engineering?*

### Keywords:

The object oriented approach of software engineering can also be applied to mixware engineering with some necessary modifications and enrichments.

**Aspect one:** static structure and dynamic behavior of objects:

The first modification is the definition of objects in the mixware context. Here we have three kinds of objects: software objects (including software connector objects), knowware objects and knowledge middleware objects. First we delineate their static structures. A software object has the same structure as it has in conventional software engineering. It mainly consists of two parts: data part and method part. A knowware object has only data part, but no method part. This corresponds to the definition of knowware that it is a knowledge module without control. A knowledge middleware object is basically a software object since knowledge middleware is a particular kind of software. It differs from a software object only in the content of the data part, which can be empty for a knowledge middleware object. Note that a software object always has a non empty data part. Figure 7 shows the difference of the three kinds of objects.

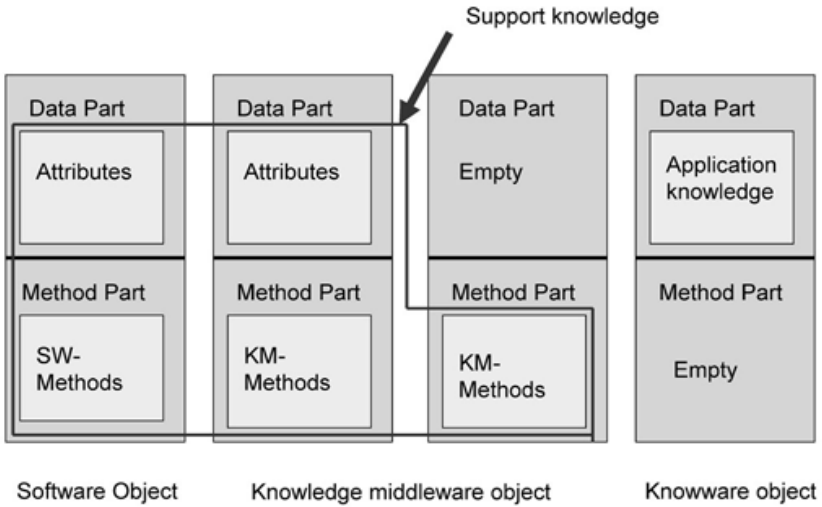

Figure 7. Static Structure of Objects

The size of a knowware object depends on the granule of knowledge it contains. A large size knowware object may be decomposed in two or more smaller objects. The meaning of doing this way is not only to increase the modularity, but is also to separate environment sensitive knowledge from environment insensitive knowledge (see discussion below). This separation makes the renewal of knowware easier and more cost effective during system evolution.

Now some words on their dynamic behavior. Let's first observe the knowware objects. It is clear that an object without method part cannot do anything actively. Moreover, such object cannot be called by other objects passively, either, since in order to be called, an object must have method part. Therefore, before a mixware program runs, each knowware object must be bound with at least one knowledge middleware object. The binding mechanism is simple and proceeds as follows: 1. For each knowledge middleware object, as many copies of it will be made as the number of knowware objects that have to be bound with it. 2. For each knowware object KO, assume $\{KMO_1, \ldots, KMO_n\}$ be the set of copies of different knowledge middleware objects that have to be bound with KO, let NKO be the object whose data part is the integration of the data parts of all $KMO_i$ with the data part of KO, and whose method part is the integration of the method parts of all $KMO_i$. Then NKO is the run time (KO + KMO) object we need. 3. A careful adjustment of inheritance relation is needed, such that each object or object class precisely inherits the same thing from their ancestors just as if they were not combined. Note that some renaming of data part items and method part items may be necessary when doing the integration. Note also that the software objects do not need any modification. After the integration is done, we obtain a new program consisting of the original set of software objects and a new set of objects that are combinations of knowware and knowledge middleware objects. Now all objects are in the conventional form and the program can start to run. Figure 8 shows the dynamic behavior of objects.

Note that there are two alternatives. The integration of knowware objects with knowledge middleware objects can be done right at the beginning when the mixware is loaded. Or it can also be done each time when the program runs. The first alternative is static. It saves run time. The second approach is dynamic. It saves space. There are also techniques for optimization. For example, two knowware objects may be merged together when the knowledge middleware objects applying to them are the same and/or when these two objects are almost always used simultaneously.

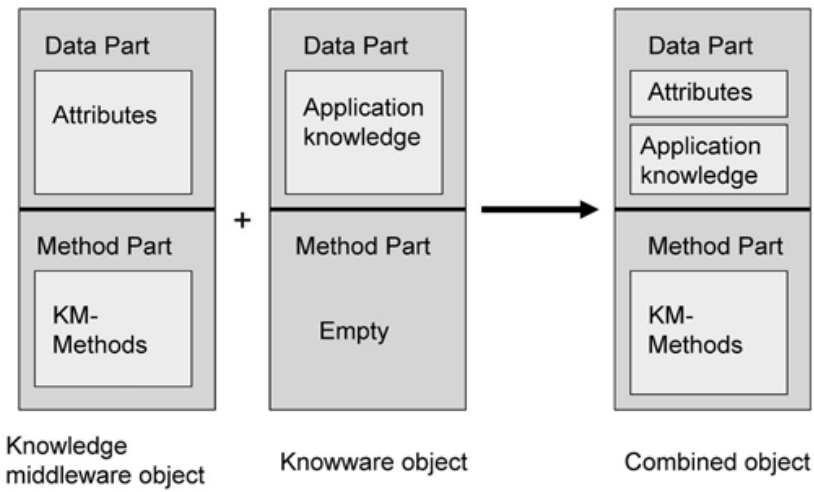

Figure 8. Dynamic Behavior of Objects

**Aspect two:** object-oriented analysis:

Given a system specification, the standard approach is to analyze the specification text and to decompose the specification in objects. In mixware engineering, the first step is almost the same as OOA of traditional software engineering. We scan the specification text and look for any entity, which can be made to object, no matter what kind of object: be it software object, knowware object or knowledge middleware object. We call them object embryos or just stem objects. The term 'stem' is borrowed from the biology where one calls those cells that are not fully differentiated as stem cells. Here we call those objects that are not yet classified as software objects, knowware objects or knowledge middleware objects as stem objects.

The second step is to differentiate the stem objects in their types. The principle is the so-called tower principle, which adopts a knowledge centered decomposition and composition strategy to determine the types of objects and their classes. According to this principle, we decompose a system requirement or a specification based on the knowledge richness and knowledge processing relevance of the system components. The richer the application knowledge contained in an object is, the more probably will this object be considered as candidate of a knowware object. Furthermore, among all knowledge intensive or knowledge processing relevant objects of a system, we distinguish those objects based on the rate of stability of the knowledge content. The more frequent an object's knowledge is subject to change (i.e. environment sensitive), the more probably will this object be considered as candidate of a knowware object. Thirdly, we classify the objects according to their interdependency of requirements. Any object will be more plausibly considered as candidate of knowware object if its knowledge requirement depends on other objects or on exterior knowledge sources. In summary, we give always priority to those system components that are (application) knowledge poor and subject to least frequent changes. We call it the tower principle because it simulates the idea of oil fractionating tower, where heavy and light oils are separated according to the richness of carbon molecules they contain, which determines the boiling point of the oil. Thus our OOA model is also called tower model.

**Aspect three:** object-oriented design:

The tower model covers also object oriented design. Remember that in an oil fractionating tower, the heavy and light oils come out in different order. Light oils come out first because they have fewer carbon molecules and therefore lower boiling points. During the OOD phase, we first determine the pure operational software objects and their connectors, which are most close to functional requirements of the system. Then we determine the knowledge needed by these software functions. Objects containing such knowledge will be made to knowware objects by removing their method parts, which will be later wrapped in form of knowledge middleware objects. The third step is to make sure that all knowledge sources needed by the knowware are located. The requirement of knowledge source depends on that of knowware. At last, we consider and reorganize all knowledge middleware objects, which are bridges between knowware objects and all other objects including knowledge sources. The procedure of generating knowledge middleware objects is as follows. The first knowledge middleware objects are made of method parts separated from knowledge rich software objects. From a knowledge rich software object we may produce one or more knowledge middleware objects, depending on how we separate the method part. Second, we may find that we need a bridge between a knowware object and another object (software, knowware, knowledge middleware, knowledge source). Then we generate a new knowledge middleware object to connect them. Third, we may find that we need more software functions to meet the user requirements. This may imply a need of new knowledge processing tools for some knowware objects. As a consequence, we have to design and implement new knowledge middleware objects. Figure 9 shows the idea of tower model. Each object at arrow head comes out earlier than that at arrow tail.

Intuitively, the tower model is a modification of the idea of the fountain model of software engineering process. According to the tower model, there is also a fountain, from which the objects are sprayed out. But the objects have types (software objects, knowware objects or knowledge middleware objects), which determine the order of objects sprayed out.

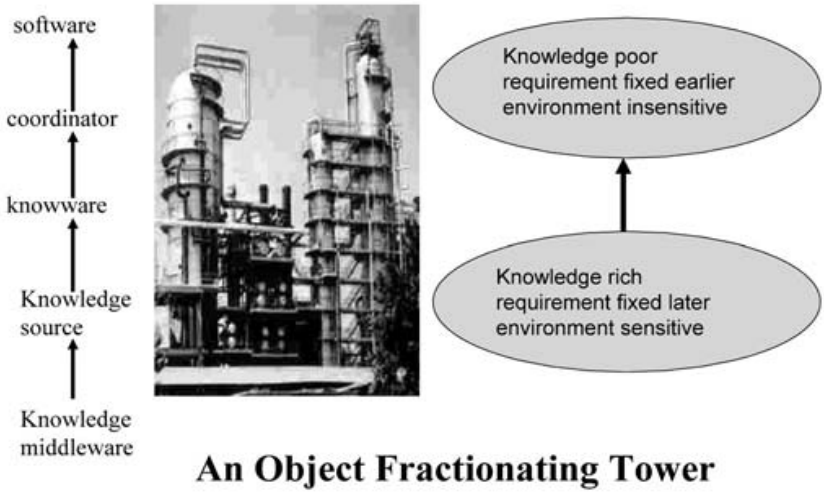

Figure 9. The Tower Model

## *What is component-oriented mixware engineering?*

### KEYWORDS:
COMPONENT ORIENTED MIXWARE ENGINEERING, SOFTWARE COMPONENT, KNOWWARE COMPONENT, GLUING MIDDLE COMPONENT, GLUING MEMBRANE, GLUING MIDDLEWARE, SOFTWARE MEMBRANE, SOFTWARE CONTAINER, KNOWLEDGE VIEW, SOFTWARE CONNECTOR, KNOWLEDGE MIDDLEWARE, PLUG-INS, XBEANS, .NET, FRACTAL, REO, PKUAS

Just as component based engineering has become a standard approach in software engineering[15], the use of component structure is also essential in mixware engineering. The major difference to a software component is that a mixware component may contain three kinds of sub-components: software components, knowware components and gluing middleware components. A gluing middleware is either a software connector or a knowledge middleware. The former is used to connect two or more software components in conventional software engineering. The latter connects a knowware component with another component, be it a software component or another knowware component, or even an exterior knowledge source. A knowware component follows the definition of knowware given at the beginning of this book. Its representation may be object oriented or not. In this context we assume it is object oriented. Besides, it is wrapped in a shell called knowledge interface. This knowledge interface is responsible for providing any information about the knowware object to easy its use by other components.

A component needs some means to "glue" its subparts together. These subparts may be objects or sub-components. Let's compare our approach with current component based software engineering techniques. There are different approaches to do that. 1. The fortified object approach. This approach keeps the object-oriented framework basically unchanged by adding some interconnecting functions to it. Examples are the plug-ins, Xbeans, .net, etc. 2. The coordinating membrane approach, which attaches to each component a membrane containing a set of interoperation controllers. One example is the Fractal approach developed at INRIA.[37] 3. The container approach, which uses software container as coordinator, e.g. the PKUAS approach undertaken by Peking University.[31] 4. The coordinating language approach, which uses an independent coordinating language to program the coordinators, e.g. the Reo language designed by Eindhoven University[32].

Our mixware approach makes use of advantages of the above mentioned approaches, but is different from them. We assign the component gluing function separately to the membranes of sub-components on the one hand, and to the independent coordinators on the other hand. We call the former gluing membranes and the latter gluing middleware. The gluing membranes are classified as software membrane, which specifies internal control of a software component, and knowledge interface, which specifies knowledge processing protocols of a knowware component. The gluing middleware are classified as software connectors, which connect only software components, and knowledge middleware, which connects a knowware component with some other components. Figure 10 illustrates the component structure of mixware.

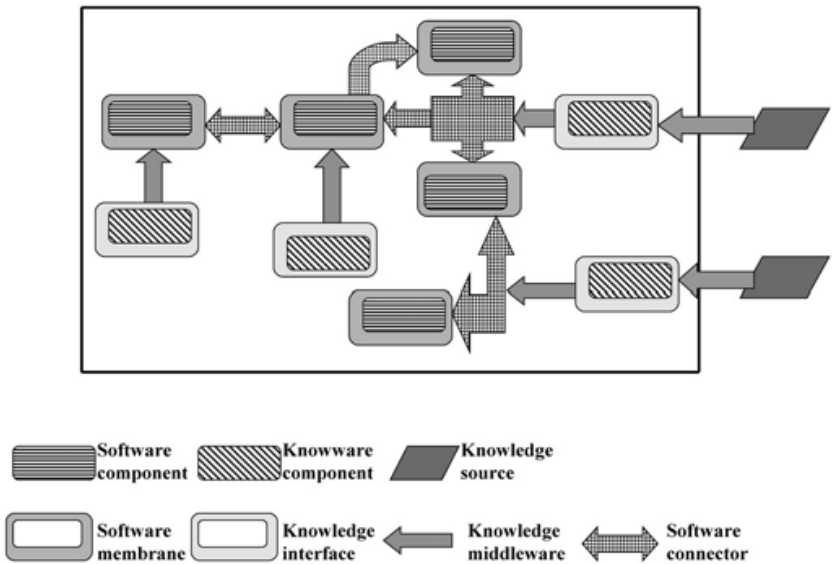

Figure 10. Component structure of Mixware

*How to do mixware modeling?*

**KEYWORDS:**
MIXWARE MODELING, UML, KUML, KNOWWARE BASED UNIVERSAL MODELING LANGUAGE, KUML DIAGRAM, KNOWLEDGE DEPENDENCY DIAGRAM, KNOWLEDGE ACTIVENESS DIAGRAM, KNOWLEDGE MIDDLEWARE DIAGRAM, KNOWLEDGE FLOW DIAGRAM

Mixware modeling is the extension of software modeling. Considering that the UML language has become a standard de facto for software modeling, mixware modeling has taken the approach of extending the UML language to KUML, which means Knowware based Universal Modeling Language. Since UML is mainly a diagram language, the extension of UML to KUML is largely implemented in augmenting and enriching the syntax and semantics of UML diagrams. Generally speaking, there are many different kinds of diagrams in UML, for example class diagram, object diagram, component diagram, package diagram, deployment diagram, use case diagram, state diagram, activity diagram, sequence diagram, timing diagram collaboration diagram, etc. We do not want to discuss their modification one by one in detail when transferring from UML to KUML, but will only delineate the main characteristics of KUML diagrams.

One of the most important modifications is to sort all classes, objects and components (they have a unified name: units) in a sequential order according to the application knowledge content they include and to its frequentness of change. More exactly, this is a sequence where each element is a group of units. (Note that a partial order is also possible. It requires a more delicate discussion) Remember that in the object oriented analysis of mixware we have used the tower model to separate knowware objects from software objects, separate application knowledge rich objects from application knowledge poor objects and separate frequent changing knowware objects from relative stable knowware objects. This order of objects will be passed over to the modeling phase and used in the construction of diagrams. The more stable and knowledge poor units (we call them light units) are in the middle layers and frequently changing and application knowledge rich units (we call them heavy units) are in the external layers.

Take the use case diagram as example. Figure 11 is a simplified use case diagram for ticket booking in UML, while figure 12 is the same use case interpreted with a KUML diagram. In figure 11, there is only one box, which includes three function units: buy ticket, ticket refund and ticket management; and one actor: the traveler. There is no difference between software and knowware.

In figure 12, the box is nested in three layers. All functional units are in the core layer. They will be programmed as operational software modules. Knowledge units are separated from function units. They will be coded as knowware. We see the two knowware units in the middle layer. They are flight time table and airline regulation. In the most outside layer there is an external knowledge source called flight change information, which is subject to most frequent change and does not belong to the category of knowware anymore. There are three actors outside of the box. They are the traveler, the airline agency and the air travel administration. These three actors operate on different functional units.

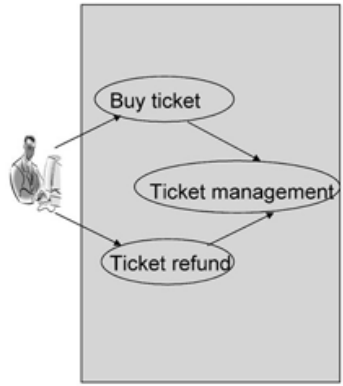

Figure 11. UML use case diagram for ticket office

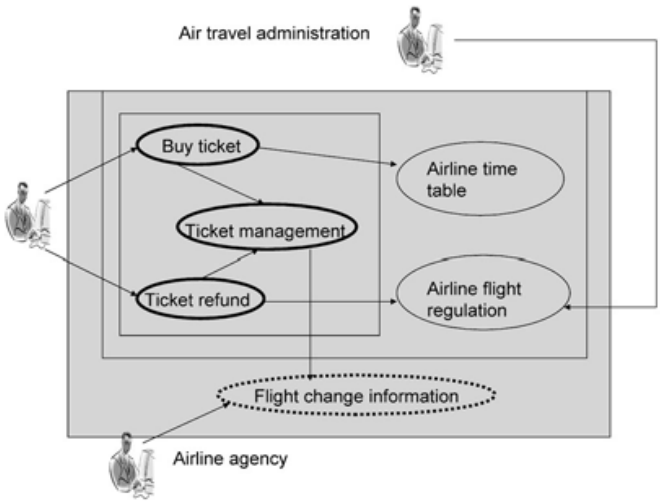

Figure 12. KUML use case diagram for ticket office

Another important difference between KUML and UML is that we have introduced new types of diagrams in KUML to describe relations between different knowledge types, between different stages of knowledge processing, and between knowware units and other units. Below are some examples:

Knowledge dependency diagram: during the OOA phase, we should distinguish three types of objects: the software objects, the knowware objects, and the knowledge middleware objects. One of the basic principles for this distinguishing is the knowledge dependency between different objects and components. Knowledge A depends on knowledge B means that B's change implies A's change. For example, the knowledge of tax law of local government depends on the knowledge of tax law of central government. The knowledge of development tendency of computer science depends on the knowledge of development tendency of its sub-branches. Therefore it is important to represent this dependency relation in a directed graph, called dependency diagram. Having such diagram established, we can then easily determine the three types of objects by using graph theoretical techniques.

Knowledge activeness diagram: another criterion for distinguishing knowware objects is the activeness of knowledge they include. Note the difference between knowledge dependency and knowledge activeness. Knowledge A is more active than knowledge B means the application of A, which plays the role of a callee, has to be supported by B, which plays the role of a caller, but not vice versa. For example, the application of X knowledge base needs support of the CAX shell. The application of knowledge contained in the 0.76 Million computer science papers needs the support of Dolphin tools, etc. Therefore it is often useful to construct a partial order of knowledge activeness regarding the current application. This is also an important criterion for classifying objects.

Knowledge middleware diagram: It models the knowledge input/output of a knowledge middleware. Knowledge middleware are the only units in a mixware complex, which operate on knowware (in a read only way) or temporary knowledge source. More exactly, they read knowledge from some knowware units (possibly using filters) and produce temporary knowledge (by possible knowledge transformation and integration) to be used by software units. Figure 13 shows the way of producing a courseware for individual use. The diagram box is nested. In the middle is the courseware specializer, which is a knowledge middleware. To its left is the basic course material, which is a knowware. To its right is the specialized courseware produced by the courseware specializer that fits the need of the individual user. Outside of the box it is the courseware user.

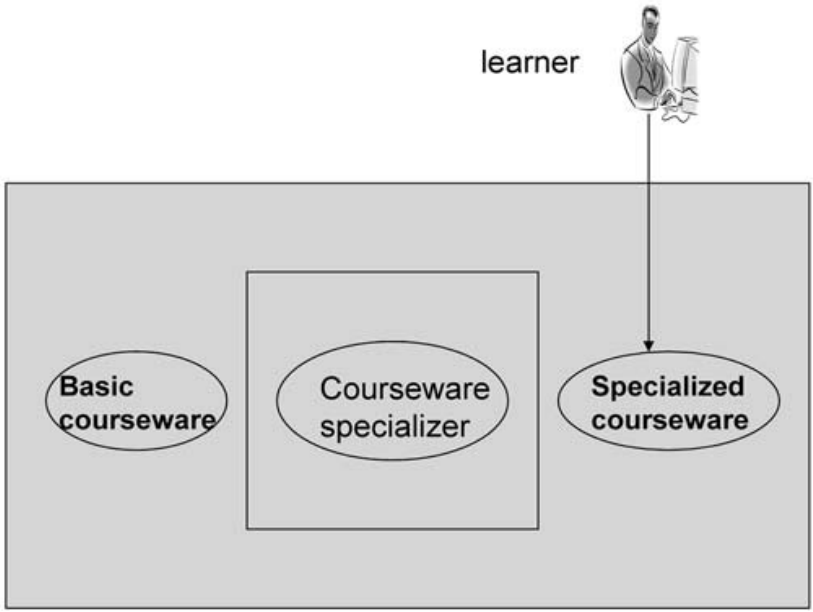

Figure 13. KUML Knowledge middleware diagram

Knowledge flow diagram: It models a complicated knowledge processing process, which can be considered as a flowchart with conditional statements and loops. But for the moment we limit us to the linear case, where the diagram represents a sequence of knowledge input/outputs. In this sense, a knowledge flow diagram can be considered as the combination of n ($\geq 1$) knowledge middleware diagrams. In the life cycle models of knowware engineering, the knowware generation process can be described by such diagrams.

*What is the basic model of knowledge service?*

**KEYWORDS:**
WEB SERVICE, KNOWLEDGE SERVICE, KNOWLEDGE SERVER, KNOWWARE BASED WEB SERVICE, THREE ELEMENTS MODEL, FOUR ELEMENTS MODEL, KNOWWARE TRANSACTION, INSTANT SUPPLY OF KNOWLEDGE, KNOWLEDGE WAREHOUSE, KNOWWARE BASED E-PUBLISHING, KNOWWARE BASED E-LEARNING

Knowledge service differs from Web service in many aspects.

First, Web service is usually based on a three elements model including a service provider, a service requester and a service agency. The service provider publishes its service at service agency, which is a central location for registering, managing and distributing services. The service agency also discovers and composes services to meet the need of service requester. The knowledge service, on the other hand, is based on a four elements model, rather than a three elements one. Besides knowledge requester, knowledge provider and knowledge server, the knowledge service model contains a fourth element: the knowledge source. This is an essential difference between knowledge service and Web service. The role of a knowledge server is not limited to detecting, composing, registering and managing knowledge services. It also generates knowware (and its half fabricates) from knowledge sources and keeps its knowledge content up to date. As we will see below, the knowledge server keeps knowledge at all fabrication layers: knowledge ore, knowledge magma, knowledge crystal, knowware and also links to external knowledge sources. The acquisition and maintenance of knowledge and evolution of knowledge crystal is done continuously 24 hours a day. This is why we have taken knowledge source as the fourth element in the model. This is also why we use the name knowledge server for the managing element of the model, rather than just call it a knowledge agency. Knowledge service is much more active and productive than a Web service agency. Figure 14 compares Web service model against knowledge service model.

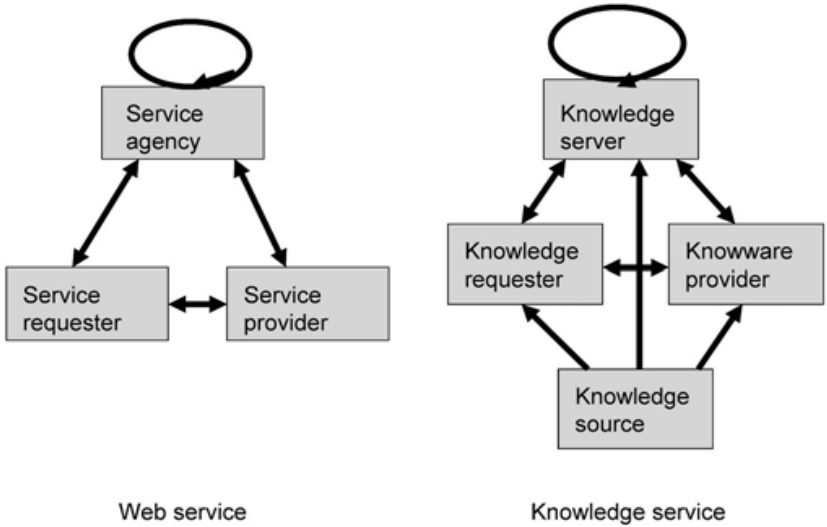

Figure 14. Knowledge service vs. Web service

Any knowledge service is impossible without the function of knowledge middleware. We list some possible ways of knowledge service:

1. Web based knowware transaction. This is something like e-commerce with knowware as the transacted commodity.

2. Instant supply of knowledge. This is done by a query and answer system operating on knowledge warehouse that is similar to the Dolphin system.

3. Knowledge source chaining. This function is like a browser. But it is better than a browser by pointing at a

minimal number of Web pages while satisfying the user requirement maximally.

4. Knowware based e-publishing. This is to meet the user requirements like: "Please write every month a report on the tendency of worldwide steel production". The accomplishment of this job would need the cooperation of several knowledge middleware, including knowledge middleware for knowledge acquisition, knowledge fusion, knowledge editing, etc.

5. Knowware based e-learning. This is to provide an entry to some e-learning platform. There are at least three levels of knowware based e-learning. 1. Provide an entry to some existing e-learning platform. 2. Reorganize an e-learning platform (in particular the knowware contained in it) to meet individual requirements. 3. Acquire knowledge needed by the user and construct a new e-learning platform with new generated knowware.

*How to construct a knowledge server?*

**KEYWORDS:**
KNOWLEDGE SERVER, KNOWLEDGE WAREHOUSE, KNOWLEDGE ORE, KNOWLEDGE MAGMA, KNOWLEDGE CRYSTAL, KNOWWARE, KNOWLEDGE MIDDLEWARE, KNOWLEDGE PROTOCOL

A knowledge server consists of four parts. The most important part is its knowledge warehouse. All fabricates and half fabricates of knowware are stored in this warehouse. They are divided in four layers: a selected set of condensed knowledge source (also called knowledge ore), a set of containers containing knowledge magma (organized sets of knowledge elements), a set of knowledge crystals (domain oriented half fabricates of knowware) and a set of knowware itself. Note that this set of knowware does not need to be an archive of knowware bodies. The major part of this set may be in form of pointers showing where the relevant knowware are. The division of the four layers is not fixed. The members of all four layers are in a steady process of upgrading: knowledge source → knowledge magma → knowledge crystal → knowware.

The second part of a knowledge server is its archive of knowledge middleware. As we have mentioned above, knowledge middleware is a class of software, which performs all jobs relating to the generation, maintenance and other operations of knowware. There is a big variety of knowledge middleware. It is not necessary and in general impossible to have all kinds of knowledge middleware on the same knowledge server. The size and content of the archive of knowledge middleware on each knowledge server may be different.

The third part of a knowledge server is its archive of knowledge protocols. Each knowledge protocol is used for translation between different knowledge representations. Here the international knowledge representation standards or standards de facto play an important role. Usually a knowledge protocol helps to translate knowledge between standard and non-standard representations, or between two different standards of knowledge representation. There are different standards available for knowledge representation, for example the traditional ones like KIF and Ontolingua, or the XML based ones like OWL, RDF and their semantic interpretation like DL. Currently, the XML based standards are preferred.

The fourth part of a knowledge server is management oriented. Its function includes service provider registration,

service requester registration, user requirement analysis, knowledge service design, knowledge content validation, knowware use authentication, etc. Its function is partly similar to the function of a Web server, and partly to that of an e-commerce service.

*How is knowledge service of use for Web service, internetware and software as a service?*

**KEYWORDS:**
WEB SERVICE, SOFTWARE AS A SERVICE, KNOWLEDGE SERVICE, KNOWWARE BASED WEB SERVICE, INTERNETWARE, INTERNET MIXWARE, SAAS

What is Web service? There are many definitions that are not identical. We take the following one, which is not most recent and formal, but more intuitive: "Web services are self-contained, self-describing, modular applications that can be published, invoked, and located across the Web. Web services perform functions, which can be anything from simple requests to complicated business processes." [11] Our question is: can Web service really do anything? The answer is: in certain sense not. A Web service cannot renew its service autonomously by renewing the knowledge contained in it, because in conventional software the knowledge is not separated from it as an independent module. This Web service must renew the relevant software as a whole, but not part of it. However, if we use mixware instead of software to construct Web service, then since in a mixware knowware and software components are separated, it is possible to only renew the knowware components, but not the software components. Therefore, knowware based Web service may work with replaceable knowledge modules of smaller size and will thus spare lots of resources.

Therefore, what we need is to replace conventional Web service with knowware based Web service. This is a uniform mechanism providing mixware service, where software service and knowledge service are both its special forms. For this new kind of service, we need another definition. Now we change the above definition on Web service as follows: Knowware based Web services are self-improving, ever-growing, modular applications that can be published, invoked, and located across the Web. Knowware based Web services perform software functions as conventional Web services with replaceable knowware parts, and provide knowledge service in various forms, from automatic knowledge acquisition, knowware generation to automatic knowledge editing and publishing.

Two points in this definition are most important. First, the new kind of service is self-improving and ever-growing. This reflects its openness character. Second, it not only provides software service, but also knowledge service.

The idea of agent based internetware, proposed by Jian lv and Hong Mei[28] aims at constructing software from software components distributed across the Web. This idea has some similarity with our crystallization model of constructing knowware from knowledge distributed across the Web. But the internetware proposal does not abstract away knowledge from software. For them, knowledge and software are still mixed together. We suggest studying internet mixware, where software and knowware are just components of the more comprehensive mixware complex.

Recently, the idea of software as a service (SaaS) has become more and more attractive, in particular to small and midsize businesses. But SaaS is actually a part of Web service and should be included in the above mentioned knowware based Web service.

It is appropriate to make a remark at this place. Above we only emphasize the use of knowledge service for Web service or internetware. But the reverse conclusion is also true. If a knowledge server receives a user requirement whose satisfaction needs some knowledge middleware that does not exist yet, then the knowledge server may call for Web service or internetware service to find, compose or generate a new knowledge middleware. This is again evidence for the necessity of introducing knowware based Web service.

*What are the challenges of knowware technology?*

**KEYWORDS:**
KNOWLEDGE ACQUISITION, E. A. FEIGENBAUM, JIM GRAY, NON CANONICAL KNOWLEDGE, KNOWLEDGE VALIDATION, KNOWLEDGE VALIDATION CENTER, KNOWWARE STANDARD

We should divide the challenges of knowware technology in two classes: those challenges that are generally applicable to any kind of knowledge processing, not necessary in connection with knowware; and those that are particularly connected with the concept of knowware.

A big challenge of the first class is knowledge acquisition, in particular, how to collect and process knowledge from an immense information space. In his position paper published in the special issue of JACM's 50$^{th}$ anniversary, Feigenbaum proposed three challenges of computer science for the first half of 21th century[10]. One of his challenges was to let computer read literature such as thick text books and build massive size knowledge base from it, such that the workload of knowledge engineering will be reduced to a tenth. Another challenge of him was to let computer search the World Wide Web and build massive size knowledge base from it, such that the workload of knowledge engineering will be reduced to a further tenth. Jim Gray proposed in the same issue[14] that we should be able to build a system such that, given a text corpus, it can answer questions about the text and summarize the text as precisely and quickly as a human expert in that field. Gray claimed further that the same thing could be done for music, images, art and cinema.

Another big challenge of the first class is the complexity associated with the processing of non canonical knowledge, reflected extensionally by its huge size, distributed nature, openness, shared, heterogeneous, multimedia or other properties, and intentionally by its ambiguity in semantics, incompleteness, inconsistency, noises, content instability, or situational/conditional context dependency, etc. Developing effective methods and tools to address those problems should be the main focus of current research in knowledge engineering.

As for the second class of challenges, the first problem we should mention is the lack of an automatic mechanism for verifying the correctness and up to date of the knowledge contained in a knowware. Note that it is more difficult to validate a knowware (service) than to validate a software (service). Since software is always oriented to the solution of a particular problem class, the simplest way of validation is to check the input/output relation of the software by considering it as a black box. But for knowware this is not enough, because knowware is universal problem solving oriented. Take a relational database as an example, which is a particular type of knowledge organization. One can check the integrity constraints of a database, but can hardly check the correctness of each piece of data in an immense database. An even more difficult problem is to check whether the knowledge contained in it is up to date. Currently, the only feasible way of guaranteeing a knowware correct and up to date is to have the authentication of an independent knowledge validation center, where all knowware providers can make an application for such authentication.

A more technical challenge is to work out standards for knowware and knowware industry. It was since long time ago that people have been working on standards of knowledge representation. But this is only part of the problem. We need many other standards, including standards of knowware-knowledge middleware interface, standards of knowledge middleware-software interface, standards of component interface, standards of knowledge service, etc. Just think how many years and how much effort we have spent (and are still spending) for determining hardware and software standards, in order to understand the heavy workload we will have to afford for these new standards.

*What kind of impact will knowware have on the development of software industry?*

**KEYWORDS:**
IMPACT OF KNOWWARE, SEPARATION OF SOFTWARE INDUSTRY, KNOWWARE INDUSTRY, IBM, DRAWBACK OF SOFTWARE ENGINEERING TECHNIQUES

The impact will be remarkable. Separation of (application) knowledge from software implies separation of knowledge package development from software package development. That means we will see the separation of the current software industry into two different, but cooperating industries: the (new) software industry that produces only system software, software middleware, software platforms and software tools, including knowledge middleware (software tools specifically dealing with knowledge processing); and knowware industry that produces knowware only. Of course, some knowware companies will also produce knowledge middleware as accompanying products of their knowware products.

According to this scenario, knowware provides a possible way to separate software developers from knowledge developers, so that software engineers only need to develop software products, while tasks of knowware development are left to various domain-specific experts. A client would buy software tools from a software developer, and buy knowware from a knowware developer. The scenario would be similar to the situation early in 1980s when a personal computer client would buy hardware from IBM and the corresponding software from compatible software developers, a win-win scenario for both hardware and software developers. We believe software and knowware would follow a similar path and become two independent and successful industries.

The introduction of knowware will also have an impact on software engineering research. The concept of software engineering was proposed in 1968 on the NATO Garmisch meeting. Since then, software engineering has become a major discipline in computer science. However, the 40 years software engineering research did not solve all its problems. In some sense, the software crisis remains to exist. That reminds us that there must be serious drawbacks in current software engineering techniques, which have not yet been discovered or attacked. We propose that one of these drawbacks is just the mixture of software and knowledge development.

*What kind of impact will knowware have on the development of knowledge industry?*

**Keywords:**

impact of knowware, edward a. feigenbaum, knowledge industry, fritz machlup, gdp growth rate, information society, knowledge based industry, knowledge intensive industry, knowledge worker, knowledge society, peter drucker, nuala beck, knowledge industry in three layers, post-industrial society, fourth industry

E.A.Feigenbaum has emphasized in his monograph about fifth generation computer that we are entering an era of "knowledge industry in which Knowledge itself will be a salable commodity like food and oil. Knowledge itself is to become the new wealth of nations"[9]. Nowadays, the concept of knowledge industry is more and more accepted by the public. But we have found that different people would have different interpretation for this terminology.

The first interpretation considers knowledge industry as an industry of producing, processing and propagating knowledge. A pioneer bearing this point of view was the American economist, Fritz Machlup, a professor of the Princeton University. He was also the first person who has proposed the concept knowledge industry[29]. According to him, the knowledge industry should include six branches. They are education, research and development, arts creation and communication, communication media, information service and information device (such as type writer). He estimated that in the United States, knowledge industry has increased with an annual rate of 10.6% in the period 1947-1958, which was about two times as high as the GDP growth rate of USA in the same period. Based on a statistics of 30 industrial branches in 1958, the contribution of knowledge industry to the global GDP of United States was about 29%. The number of people engaging in knowledge industry of USA made a 32% portion of the total number of employees in 1959. These statistics made Prof. Machlup believe that the United States has already entered the information society[29].

As a matter of fact, the knowledge industry concept of Machlup is a very general one. It is a mix of information industry (IT industry) and knowledge industry (KT). According to his definition, even telephone devices and type writers should be part of knowledge industry. But this is not what we want.

The second interpretation considers knowledge industry as "knowledge based industry". A synonym for it is "knowledge intensive industry". In these industries it is the knowledge, but not the machines or other production means, which plays a central role. When Peter Drucker studied the concept of knowledge society, he defined knowledge worker as the group of people who have mastered professional knowledge (including skill) of some domain, or experts in some particular fields. They are holding decisive positions in these industrial branches[6][7][8]. Also Feigenbaum said that the knowledge industry "creates value by transforming the brainpower of knowledge workers, with little consumption of energy and raw materials"[9]. Nuala Beck pointed out that the proportion of knowledge workers in an industry determines the degree to which extent it can be considered as knowledge industry[1].

However, it is possible to have a third interpretation, which considers knowledge industry as an industry producing knowledge (with help of computer, internet, or any other device that can process information in a massive way). More generally, we can divide knowledge industry in three nesting layers. The core layer is the industry producing knowware; the middle layer is the industry producing knowledge in general (by machine learning, data mining or what ever tech-niques); the most outside layer is the industry producing, processing and propagating knowledge, which is similar to the definition given by Machlup. But there are important differences. The first thing is that this definition distinguishes knowledge from information. For example, the news published in various media is information, but not knowledge. Secondly, our definition distinguishes activity of knowledge production from activity of knowledge propagation. Education is an example of the latter. Publishing companies and television broadcasting are other examples. Thirdly, this definition distinguishes production of knowledge itself from production of mechanical devices, which may be of use in the former activity. For example, computer industry is not a knowledge industry although it helps people to produce knowledge. Lastly, our definition distinguishes knowledge production having impact on social economy development from knowledge production of academic importance only. For example, we consider patents but not papers as knowledge products, no matter how much knowledge the papers contain. Besides, we don't think that the knowledge based industry should belong to the definition of knowledge industry in this sense, because knowledge is only a means, but not a product in these industries. With the development of science and technology, the proportion of know-how used in all industries will be more and more, but this would not say that all these industries will become knowledge industry. They will remain steel industry, aircraft industry, food industry, and so on.

We are now interested in finding the position of knowledge industry among various other industries. Usually people divide the national economy in three categories: the first industry, which is agriculture (including forestry, fishery and animal husbandry); the second industry, which is meant as enterprises of product manufacturing; and the third industry, which includes commerce, transportation, social service etc. So, to which category should knowledge industry belong?

Although the existence of a knowledge industry is an objective reality that has been recognized by the public since long time, still there are only few people who have answered this problem directly. One of the answers brought knowledge industry in tight connection with knowledge service. Daniel Bell pointed out that the rise of knowledge industry is characteristic for post-industrial societies. According to him, in a post-industrial society, the main function of the society has transferred from product manufacturing to a service economy. Theoretical knowledge, technique and information become the principal form of commodity[2]. If we accept this point of view, then knowledge industry should form a part of the third industry.

But this point of view is not in accordance with a comment of Peter Drucker, who has been discussing the role of knowledge in a future society extensively. He thought that knowledge will play an important key role in the future society, which he called as knowledge society. We are entering this knowledge society where the knowledge workers will take a leading position. Maybe they will not occupy the majority of the population, but the role they play is decisive. Everyone who contributes to a modern organization to increase its productivity with his/her particular knowledge is actually an executive officer of this organization[6][7][8]. This shows that the role of knowledge industry goes actually beyond the concept of third industry. Furthermore, knowledge industry is actually the "energy industry" of all industries. Without new knowledge, the progress of any industry is impossible.

Therefore, we reached to the conclusion that we should consider knowledge industry as the fourth industry, which is beyond the traditional three industry point of view. In a world of strong competition, the development of fourth industry is much more important than all other industries. It may determine the overall power of any country.

*Which kind of impact will knowware have on the development of IT industry in general?*

**KEYWORDS:**
NEXT GENERATION IT, INTELLECTUAL PROPERTY, THREE DIFFERENT INDUSTRIES, THREE UNDERPINNINGS OF IT, THIRD LIBERATION OF IT

Knowware is suggested as an important step towards the next generation IT and its wide applications. It is also considered as a natural development in IT after software and hardware, and should be as important as software and hardware.

Over a long period, we have mixed software development with knowledge development, software with the knowledge it provided, and particularly, the intellectual properties of software with the intellectual properties of the knowledge it used. As a result, a software developer also has to play the role of a knowledge developer, while real experts who have the domain-specific knowledge are excluded from development teams since they usually do not have the knowledge of software development. Based on the current practice of software engineering, the lack of common knowledge backgrounds between clients and software engineers will be a serious problem and will last for a long time. This is another important reason that we want to use the format of knowware to separate the knowledge component from software, and make knowware and software as two different research topics and salable commodities. A direct implication of this separation is: hardware, software, knowware become three different disciplines as well as three different industries. They should be three equally important underpinnings of the IT industry.

Historically, computer industry has been treated as part of electronics in the beginning; and the formation of an independent computer hardware industry can be considered as the first liberation of information technology from electronics. Similarly, a software package was always made to support some particular machines in the early stages of computer development. The birth of software concept and the independence of computer software development from particular hardware can be considered as the second liberation of information technology. In this sense, we suggest to consider the emergence of knowware and its independence from software as the third liberation of information technology.

*Have been the name "knowware" and related concepts used sometime before or somewhere else?*

#### KEYWORDS:

KNOWWARE, KNOWLEDGE MIDDLEWARE, KNOWLEDGE PUMP, KNOWLEDGE CRYSTAL, METALOG, ORM-ML, KNOWLEDGE SEA, KNOWLEDGE SOUP, FOURTH INDUSTRY, XEROX, JOHN F. SOWA, IBM, ALFRED SPECTOR, IKUIJIRO NONAKA AND HIROTAKA TAKEUCHI, BARRY BOEHM, SPIRAL MODEL, TACIT KNOWLEDGE, EXPLICIT KNOWLEDGE, INDIVIDUAL KNOWLEDGE, SOCIAL KNOWLEDGE, INTERNALIZATION, SOCIALIZATION, EXTERNALIZATION, COMBINATION, FOURTH INDUSTRY

After having coined the concept knowware and having published our first results of studying knowware, we have found that some of the concepts used by us had been used somewhere else for other purposes. However, after a careful check we found that these concepts, sometimes used in literature earlier than we did, do not mean exactly the same thing as we do. We mention some of them in the following.

The name of knowware or knowledge ware had been used by some company to denote some of their software products. Some companies even called their advice or suggestions to customers as knowware. Actually they were giving the name "knowware" to some technical knowledge that is usually called know-how.

The name of knowledge pump had been used by Xerox to paten one of their products[12], while in our approach it is used to name a class of techniques of knowledge acquisition and tools of implementing these techniques.

Sowa had mentioned knowledge crystals too[35], but mainly from the aspect of cognition science. Here we have defined this crystallization process from an engineering aspect and used it to model the knowware development process. Sowa has also invented the concept of knowledge soup to emphasize the fuzziness and chaos of human knowledge. Different from him, we use the concept of knowledge sea to emphasize the immense size of knowledge and its complicated structure.

The term pseudo natural language has been also used in some other literature. But to the best of our knowledge, we are the first who has used this term in published papers[19][20][18]. One recent example of using this term is the Metalog system[30], which is a query/logical system to easy reasoning on the Web. The underlying logical model of Metalog is MLL (Metalog Logical Level), which is an extension of the RDF semantics. It has a user interface of querying in pseudo natural language. Another example is the markup language ORM-ML for representing ORM (Object Role Modeling) schemas[34]. They have developed a verbalized style sheet for ORM-ML documents allowing presenting facts and rules in pseudo natural language sentences, which are in fact translations of a graphical representation.

Compared to the work reported in literature, our PNLU technique has the following characteristics:

1. Our PNL is a wide spectrum language family from natural languages till formal languages. Its layers form a partial order structure and can be defined arbitrary depending on user applications.

2. Each member of PNL has a key structure that is syntactically, semantically and pragmatically defined.

3. PNLU is systematically applied to large scale knowledge acquisition from written texts, domain experts, simple users and large information space such as WWW.

Also the name knowledge middleware is not new. Alfred Spector had addressed issues related to knowledge middleware in many places in his keynote report at WWW 2002[36]. In his opinion, knowledge middleware is a kind of middleware used to acquire knowledge massively, especially through text mining, while in our terminology knowledge middleware denotes any software or program helping the interoperation of knowware with any one of the following: software, hardware, human user, knowledge source, or another knowware in any kind of knowledge processing.

In 1995, Nonaka and Takeuchi proposed a model of human knowledge accumulation and evolution[33]. The model is called spiral model, which describes the process of knowledge transformation between tacit knowledge and explicit knowledge through a spiral of internalization, socialization, externalization and combination, which characterizes the formation and transformation of tacit and explicit knowledge, as well as individual and social knowledge.

It circles the loop:

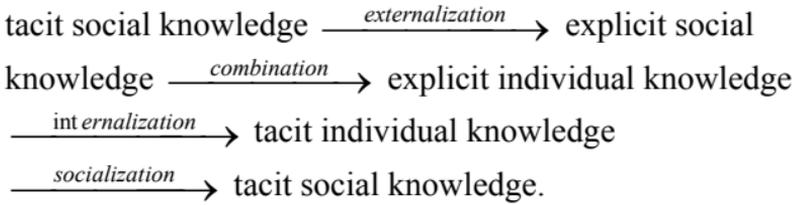

tacit social knowledge $\xrightarrow{externalization}$ explicit social knowledge $\xrightarrow{combination}$ explicit individual knowledge $\xrightarrow{internalization}$ tacit individual knowledge $\xrightarrow{socialization}$ tacit social knowledge.

We adapt this model to describe the life cycle of knowware engineering, in particular for the case when knowledge is acquired mainly by a process of experience accumulation. As a matter of fact, the concept of spiral model went back to a time much earlier when Barry Boehm published his model of software economy[3]. The model of Boehm, though, can also be considered as a knowledge evolution model that delineates the process of evolution of software engineer's knowledge about a software project along with the progress of this project.

Also the concept of fourth industry had been used to mean something completely different than what we mean. Some examples are: education, banking industry and solid waste reuse. According to our point of view, all these definitions are questionable. Even if they could be called as fourth industry, none of them could have the same importance as the knowledge industry.

## Acknowledgement

I would like to thank Fei-Yue Wang, Daniel Zeng, Zhi Jin, Hong Mei, Jian Lv, Lin Liu, Xiaoguang Yang and Yong Shi for their valuable encouragement and discussion on Knowware. This work is supported in part by the NSFC Major Research Program 60496324, NSFC project 60603002 and 973 project 2002CB312004.

# REFERENCES

[1] N. Beck, Shifting Gears - Thriving in the New Economy, HarperCollins, Toronto, 1992.

[2] D. Bell, The coming of post-industrial society: A Venture in Social Forecasting, Penguin, 1973.

[3] B. Boehm, A Spiral Model for Software Development and Enhancement, Computer, Vol. 21, No. 5, May 1988, pp. 61-72.

[4] R.A. Brooks, Intelligence without reason, Proc. of IJCAI '91, pp. 569-604.

[5] C. Cao et al., Progress of national knowledge infrastructure, J. Computer Science and Technology, Vol. 17, pp. 523-534. 2002.

[6] P. F. Drucker, The effective executive, HarperCollins Publishers, 1966.

[7] P. F. Drucker, The age of discontinuity, Transaction Publishers, 1969.

[8] F. Drucker, Post Capitalist Society, Harper Business Publishers, New York, 1994.

[9] E.A. Feigenbaum and P. McCorduck, The Fifth Generation: Artificial Intelligence and Japan's Challenge to the World, Addison-Wesley, 1983.

[10] E.A. Feigenbaum, Some challenges and grand challenges for computational intelligence, JACM, Jan. 2003, pp. 32-40.

[11] C. Ferris & J. Farrell, What are Web services? CACM, 46(6), pp. 31-35, 2003.

[12] N. Glance et. al., Knowledge Pump: Supporting the Flow and Use of Knowledge, in Information Technology for Knowledge Management (eds. U. Borghoff et. al.), Ch3, Springer-Verlag, 1998.

[13] Google Corporate Information, Google milestones http//www.google.com/corporate/history.html

[14] J.Gray, What next?: A dozen information-technology research goals, JACM, Jan. 2003, pp. 41-57.

[15] J. Kerr et. al., Inside RAD, McGraw-Hill, 1994.

[16] D. B. Lenat, CYC: A Large-Scale Investment in Knowledge Infrastructure, CACM, V. 38, pp. 32-38, 1995.

[17] R. Lu and C. Cao, Towards Knowledge Acquisition from Domain Books, Current Trends in Knowledge Acquisition, pp. 289-301, IOC, Amsterdam, 1990.

[18] R. Lu, C. Cao, Y. Chen, W. Mao, W. Chen, Z. Han, The PLNU approach to automatic generation of ICAI systems, Science in China, series A, Vol. 38, supplement, pp. 1-11, 1995.

[19] R. Lu, Automatic Knowledge Acquisition by Understanding Pseudo-Natural Languages, Theory and Praxis of Machine Learning, Dagstuhl Seminar Report 91 (9426), pp. 11-12, 1994.6.

[20] R. Lu, Z. Jin and R. Wan, Requirement specification in pseudo-natural language in PROMIS, proc. of 19[th] COMPSAC, pp. 96-101, 1995.

[21] R. Lu and Z. Jin, Domain Modeling Based Software Engineering – A Formal Approach, Kluwer Publishing Co., 2000.

[22] R. Lu, C. Shi, S. Zhang, X. Mao, J. Xu, P. Yang, L. Fan, Agent-oriented commonsense knowledge base, Science in China, E, (43)6, pp. 641-652, 2000.

[23] R. Lu, From hardware to software to knowware: IT' s third liberation? IEEE Intelligent systems, Mar/Apr, 2005. pp.82-85.

[24] R. Lu, D. Zeng, F. Wang, AI reserch in China, 50 years down the road, IEEE Intelligent systems, May/June, 2006. pp. 91-93.

[25] R. Lu, Z. Jin, Beyond knowledge engineering, JCST, Sept. 2006.

[26] R. Lu, Towards a software/knowware co-engineering, invited talk, in J. Lang, F. Lin, J. Wang (eds.), KSEM06, LNCS 4092, pp. 23-32, 2006.

[27] R. Lu, Knowware, knowware engineering and software/ knowware co-engineering, invited talk, ICCS '07, Beijing, May, 2007.

[28] J. Lv et. al. Research and Development of Internetware, Science in China E, 36(10):1037-1080, 2006.

[29] F. Machlup, The Production and Distribution of Knowledge in the United States, Princeton University Press, 1962.

[30] M. Marchiori, Towards a People's Web: Metalog, 2004 http://www.w3.org/People/Massimo/papers/2004/wi2004.pdf

[31] H. Mei, G. Huang, PKUAS: An Architecture-based Reflective Component Operating Platform, invited paper, 10$^{th}$ IEEE International Workshop on Future Trends of Distributed Computing Systems (FTDCS), pp. 163-169, 2004.

[32] M.R. Mousavi, M. Sirjani, F. Arbab, Formal Semantics and Analysis of Component Connectors in Reo, FOCLASA'05, Electronic Notes in Theoretical Computer Science, Elsevier Science B.V., August 2005.

[33] I. Nonaka & H. Takeuchi, The Knowledge Creating Company: How Japanese Companies Create Dynamics of Innovation, Oxford University Press, 1995.

[34] OASIS, STARLab ORM Markup Language (ORM-ML), 2002 http://xml.coverpages.org/ORM-200206.pdf

[35] J.F. Sowa, Representing Knowledge Soup in Language and Logic, Conference on Knowledge and Logic, Darmstadt, 2002.

[36] A. Spector, Architecting Knowledge Middleware, WWW2002, Hawaii, 2002.

[37] J-B Stefani, Fractal, A component model for manageable distributed systems, http://fractal.objectweb.org

[38] A. Zhou, Annual report of NSFC major research program 60496325.

[39] H. Zhuge, Knowledge Grid, World Scientific, 2004.

# LIST OF KEYWORDS

(Some keywords appear duplicated in order to facilitate the search)

Creative Commons Legal Code
Attribution-NonCommercial 3.0 Unported

**License**

      not be considered an Adaptation (as defined above) for the purposes of this License.

c) "**Distribute**" means to make available to the public the original and copies of the Work or Adaptation, as appropriate, through sale or other transfer of ownership.

d) "**Licensor**" means the individual, individuals, entity or entities that offer(s) the Work under the terms of this License.

e) "**Original Author**" means, in the case of a literary or artistic work, the individual, individuals, entity or entities who created the Work or if no individual or entity can be identified, the publisher; and in addition (i) in the case of a performance the actors, singers, musicians, dancers, and other persons who act, sing, deliver, declaim, play in, interpret or otherwise perform literary or artistic works or expressions of folklore; (ii) in the case of a phonogram the producer being the person or legal entity who first fixes the sounds of a performance or other sounds; and, (iii) in the case of broadcasts, the organization that transmits the broadcast.

f) "**Work**" means the literary and/or artistic work offered under the terms of this License including without limitation any production in the literary, scientific and artistic domain, whatever may be the mode or form of its expression including digital form, such as a book, pamphlet and other writing; a lecture, address, sermon or other work of the same nature; a dramatic or dramatico-musical work; a choreographic work or entertainment in dumb show; a musical composition with or without words; a cinematographic work to which are assimilated works expressed by a process analogous to cinematography; a work of drawing, painting, architecture, sculpture, engraving or lithography; a photographic work to which are assimilated works expressed by a process analogous to photography; a work of applied art; an illustration, map, plan, sketch or three-dimensional work relative to geography, topography, architecture or science; a performance; a broadcast; a phonogram; a compilation of data to the extent it is protected as a copyrightable work; or a work performed by a variety or circus performer to the extent it is not otherwise considered a literary or artistic work.

g) "**You**" means an individual or entity exercising rights under this License who has not previously violated the terms of this License with respect to the Work, or who has received express permission from the Licensor to exercise rights under this License despite a previous violation.

h) "**Publicly Perform**" means to perform public recitations of the Work and to communicate to the public those public recitations, by any means or process, including by wire or wireless means or public digital performances; to make available to the public Works in such a way that members of the public may access these Works from a place and at a place individually chosen by them; to perform the Work to the public by any means or process and the communication to the public of the performances of the Work, including by public digital performance; to broadcast and rebroadcast the Work by any means including signs, sounds or images.

- i) "**Reproduce**" means to make copies of the Work by any means including without limitation by sound or visual recordings and the right of fixation and reproducing fixations of the Work, including storage of a protected performance or phonogram in digital form or other electronic medium.

**2. Fair Dealing Rights.** Nothing in this License is intended to reduce, limit, or restrict any uses free from copyright or rights arising from limitations or exceptions that are provided for in connection with the copyright protection under copyright law or other applicable laws.

**3. License Grant.** Subject to the terms and conditions of this License, Licensor hereby grants You a worldwide, royalty-free, non-exclusive, perpetual (for the duration of the applicable copyright) license to exercise the rights in the Work as stated below:
- a) to Reproduce the Work, to incorporate the Work into one or more Collections, and to Reproduce the Work as incorporated in the Collections;
- b) to create and Reproduce Adaptations provided that any such Adaptation, including any translation in any medium, takes reasonable steps to clearly label, demarcate or otherwise identify that changes were made to the original Work. For example, a translation could be marked "The original work was translated from English to Spanish," or a modification could indicate "The original work has been modified.";
- c) to Distribute and Publicly Perform the Work including as incorporated in Collections; and,
- d) to Distribute and Publicly Perform Adaptations.

The above rights may be exercised in all media and formats whether now known or hereafter devised. The above rights include the right to make such modifications as are technically necessary to exercise the rights in other media and formats. Subject to Section 8(f), all rights not expressly granted by Licensor are hereby reserved, including but not limited to the rights set forth in Section 4(d).

**4. Restrictions.** The license granted in Section 3 above is expressly made subject to and limited by the following restrictions:
- a) You may Distribute or Publicly Perform the Work only under the terms of this License. You must include a copy of, or the Uniform Resource Identifier (URI) for, this License with every copy of the Work You Distribute or Publicly Perform. You may not offer or impose any terms on the Work that restrict the terms of this License or the ability of the recipient of the Work to exercise the rights granted to that recipient under the terms of the License. You may not sublicense the Work. You must keep intact all notices that refer to this License and to the disclaimer of warranties with every copy of the Work You Distribute or Publicly Perform. When You Distribute or Publicly Perform the Work, You may not impose any effective technological measures on the Work that restrict the ability of a recipient of the Work from You to exercise the rights granted to that recipient under the terms of the License. This Section 4(a) applies to the Work as incorporated in a Collection, but this does not require the Collection apart from the Work itself to be made subject to the terms of this License. If You create a Collection, upon notice from any Licensor You must, to the extent practicable, remove from the Collection any credit as required by Section

    4(c), as requested. If You create an Adaptation, upon notice from any Licensor You must, to the extent practicable, remove from the Adaptation any credit as required by Section 4(c), as requested.

b) You may not exercise any of the rights granted to You in Section 3 above in any manner that is primarily intended for or directed toward commercial advantage or private monetary compensation. The exchange of the Work for other copyrighted works by means of digital file-sharing or otherwise shall not be considered to be intended for or directed toward commercial advantage or private monetary compensation, provided there is no payment of any monetary compensation in connection with the exchange of copyrighted works.

c) If You Distribute, or Publicly Perform the Work or any Adaptations or Collections, You must, unless a request has been made pursuant to Section 4(a), keep intact all copyright notices for the Work and provide, reasonable to the medium or means You are utilizing: (i) the name of the Original Author (or pseudonym, if applicable) if supplied, and/or if the Original Author and/or Licensor designate another party or parties (e.g., a sponsor institute, publishing entity, journal) for attribution ("Attribution Parties") in Licensor's copyright notice, terms of service or by other reasonable means, the name of such party or parties; (ii) the title of the Work if supplied; (iii) to the extent reasonably practicable, the URI, if any, that Licensor specifies to be associated with the Work, unless such URI does not refer to the copyright notice or licensing information for the Work; and, (iv) consistent with Section 3(b), in the case of an Adaptation, a credit identifying the use of the Work in the Adaptation (e.g., "French translation of the Work by Original Author," or "Screenplay based on original Work by Original Author"). The credit required by this Section 4(c) may be implemented in any reasonable manner; provided, however, that in the case of a Adaptation or Collection, at a minimum such credit will appear, if a credit for all contributing authors of the Adaptation or Collection appears, then as part of these credits and in a manner at least as prominent as the credits for the other contributing authors. For the avoidance of doubt, You may only use the credit required by this Section for the purpose of attribution in the manner set out above and, by exercising Your rights under this License, You may not implicitly or explicitly assert or imply any connection with, sponsorship or endorsement by the Original Author, Licensor and/or Attribution Parties, as appropriate, of You or Your use of the Work, without the separate, express prior written permission of the Original Author, Licensor and/or Attribution Parties.

d) For the avoidance of doubt:
  i. i.Non-waivable Compulsory License Schemes. In those jurisdictions in which the right to collect royalties through any statutory or compulsory licensing scheme cannot be waived, the Licensor reserves the exclusive right to collect such royalties for any exercise by You of the rights granted under this License;
  ii. e.Waivable Compulsory License Schemes. In those jurisdictions in which the right to collect royalties through any statutory or compulsory licensing scheme can be waived, the Licensor reserves the exclusive right to collect such royalties for any exercise by You of the rights

       granted under this License if Your exercise of such rights is for a purpose or use which is otherwise than noncommercial as permitted under Section 4(b) and otherwise waives the right to collect royalties through any statutory or compulsory licensing scheme; and,

   iii. f.Voluntary License Schemes. The Licensor reserves the right to collect royalties, whether individually or, in the event that the Licensor is a member of a collecting society that administers voluntary licensing schemes, via that society, from any exercise by You of the rights granted under this License that is for a purpose or use which is otherwise than noncommercial as permitted under Section 4(c).

e) Except as otherwise agreed in writing by the Licensor or as may be otherwise permitted by applicable law, if You Reproduce, Distribute or Publicly Perform the Work either by itself or as part of any Adaptations or Collections, You must not distort, mutilate, modify or take other derogatory action in relation to the Work which would be prejudicial to the Original Author's honor or reputation. Licensor agrees that in those jurisdictions (e.g. Japan), in which any exercise of the right granted in Section 3(b) of this License (the right to make Adaptations) would be deemed to be a distortion, mutilation, modification or other derogatory action prejudicial to the Original Author's honor and reputation, the Licensor will waive or not assert, as appropriate, this Section, to the fullest extent permitted by the applicable national law, to enable You to reasonably exercise Your right under Section 3(b) of this License (right to make Adaptations) but not otherwise.

**5. Representations, Warranties and Disclaimer**

UNLESS OTHERWISE MUTUALLY AGREED TO BY THE PARTIES IN WRITING, LICENSOR OFFERS THE WORK AS-IS AND MAKES NO REPRESENTATIONS OR WARRANTIES OF ANY KIND CONCERNING THE WORK, EXPRESS, IMPLIED, STATUTORY OR OTHERWISE, INCLUDING, WITHOUT LIMITATION, WARRANTIES OF TITLE, MERCHANTIBILITY, FITNESS FOR A PARTICULAR PURPOSE, NONINFRINGEMENT, OR THE ABSENCE OF LATENT OR OTHER DEFECTS, ACCURACY, OR THE PRESENCE OF ABSENCE OF ERRORS, WHETHER OR NOT DISCOVERABLE. SOME JURISDICTIONS DO NOT ALLOW THE EXCLUSION OF IMPLIED WARRANTIES, SO SUCH EXCLUSION MAY NOT APPLY TO YOU.

**6. Limitation on Liability.** EXCEPT TO THE EXTENT REQUIRED BY APPLICABLE LAW, IN NO EVENT WILL LICENSOR BE LIABLE TO YOU ON ANY LEGAL THEORY FOR ANY SPECIAL, INCIDENTAL, CONSEQUENTIAL, PUNITIVE OR EXEMPLARY DAMAGES ARISING OUT OF THIS LICENSE OR THE USE OF THE WORK, EVEN IF LICENSOR HAS BEEN ADVISED OF THE POSSIBILITY OF SUCH DAMAGES.

**7. Termination**

a) This License and the rights granted hereunder will terminate automatically upon any breach by You of the terms of this License. Individuals or entities who have received Adaptations or Collections from You under this License, however, will not have their licenses terminated provided such individuals or entities remain in full compliance with those licenses. Sections 1, 2, 5, 6, 7, and 8 will survive any termination of this License.

b) Subject to the above terms and conditions, the license granted here is perpetual (for the duration of the applicable copyright in the Work). Notwithstanding the above, Licensor reserves the right to release the Work under different license terms or to stop distributing the Work at any time; provided, however that any such election will not serve to withdraw this License (or any other license that has been, or is required to be, granted under the terms of this License), and this License will continue in full force and effect unless terminated as stated above.

## 8. Miscellaneous

a) Each time You Distribute or Publicly Perform the Work or a Collection, the Licensor offers to the recipient a license to the Work on the same terms and conditions as the license granted to You under this License.
b) Each time You Distribute or Publicly Perform an Adaptation, Licensor offers to the recipient a license to the original Work on the same terms and conditions as the license granted to You under this License.
c) If any provision of this License is invalid or unenforceable under applicable law, it shall not affect the validity or enforceability of the remainder of the terms of this License, and without further action by the parties to this agreement, such provision shall be reformed to the minimum extent necessary to make such provision valid and enforceable.
d) No term or provision of this License shall be deemed waived and no breach consented to unless such waiver or consent shall be in writing and signed by the party to be charged with such waiver or consent.
e) This License constitutes the entire agreement between the parties with respect to the Work licensed here. There are no understandings, agreements or representations with respect to the Work not specified here. Licensor shall not be bound by any additional provisions that may appear in any communication from You. This License may not be modified without the mutual written agreement of the Licensor and You.
f) The rights granted under, and the subject matter referenced, in this License were drafted utilizing the terminology of the Berne Convention for the Protection of Literary and Artistic Works (as amended on September 28, 1979), the Rome Convention of 1961, the WIPO Copyright Treaty of 1996, the WIPO Performances and Phonograms Treaty of 1996 and the Universal Copyright Convention (as revised on July 24, 1971). These rights and subject matter take effect in the relevant jurisdiction in which the License terms are sought to be enforced according to the corresponding provisions of the implementation of those treaty provisions in the applicable national law. If the standard suite of rights granted under applicable copyright law includes additional rights not granted under this License, such additional rights are deemed to be included in the License; this License is not intended to restrict the license of any rights under applicable law.

**Creative Commons Notice**

Creative Commons is not a party to this License, and makes no warranty whatsoever in connection with the Work. Creative Commons will not be liable to You or any party on any legal theory for any damages whatsoever, including without limitation any general, special, incidental or consequential damages arising in connection to this license. Notwithstanding the foregoing two (2) sentences, if Creative Commons has expressly identified itself as the Licensor hereunder, it shall have all rights and obligations of Licensor.

Except for the limited purpose of indicating to the public that the Work is licensed under the CCPL, Creative Commons does not authorize the use by either party of the trademark "Creative Commons" or any related trademark or logo of Creative Commons without the prior written consent of Creative Commons. Any permitted use will be in compliance with Creative Commons' then-current trademark usage guidelines, as may be published on its website or otherwise made available upon request from time to time. For the avoidance of doubt, this trademark restriction does not form part of the License.

Creative Commons may be contacted at http://creativecommons.org/.

**Publishing studies** series

This book proposes to separate knowledge from software and to make it a commodity that is called knowware.
The architecture, representation and function of Knowware are discussed. The principles of knowware engineering and its three life cycle models: furnace model, crystallization model and spiral model are proposed and analyzed. Techniques of software/knowware co-engineering are introduced. A software component whose knowledge is replaced by knowware is called mixware. An object and component oriented development schema of mixware is introduced. In particular, the tower model and ladder model for mixware development are proposed and discussed.
Finally, knowledge service and knowware based Web service are introduced and compared with Web service. In summary, knowware, software and hardware should be considered as three equally important underpinnings of IT industry.

Ruqian Lu is a professor of computer science of the Institute of Mathematics, Academy of Mathematics and System Sciences. He is a fellow of Chinese Academy of Sciences. His research interests include artificial intelligence, knowledge engineering and knowledge based software engineering. He has published more than 100 papers and 10 books. He has won two first class awards from the Academia Sinica and a National second class prize from the Ministry of Science and Technology. He has also won the sixth Hua Loo-keng Mathematics Prize.

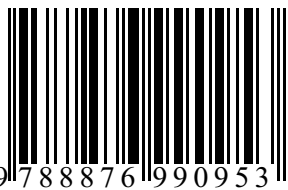